\documentclass[aps,prd,twocolumn,showpacs,preprintnumbers,amsmath,amssymb]{revtex4}
\usepackage{graphicx}% Include figure files
\usepackage{dcolumn}% Align table columns on decimal point
\usepackage{bm}% bold math
\usepackage{slashed}

\usepackage{subfigure}

\def\mmu{$\mu^{-}$ -- $\mu^{+}$}
\def\ttau{$\tau^{-}$ -- $\tau^{+}$}
\def\e+{e$^{+}$}

\begin{document}

\preprint{}

\title{On the possibility of observable signatures of leptonium  from astrophysical sources}%
 
\author{S. C. Ellis}
\email{sellis@aao.gov.au}
\affiliation{%
Australian Astronomical Observatory, 105 Delhi Rd., North Ryde, NSW 2113, Australia\\
Sydney Institute for Astronomy, School of Physics, University of Sydney, NSW 2006, Australia
}%

\author{Joss Bland-Hawthorn}
\email{jbh@physics.usyd.edu.au}
\affiliation{%
Sydney Institute for Astronomy, School of Physics, University of Sydney, NSW 2006, Australia
}%

\date{\today}% It is always \today, today,

\begin{abstract}
The formation of Ps in our Galaxy is well measured, and has led to important and unanswered questions on the origin of the positrons.  In principle it should be possible to form analogous systems from $\mu$ and $\tau$ leptons, viz.\ true muonium and true tauonium.  However the probability of formation for these systems is greatly reduced due to the intrinsically short lifetimes of the $\mu$ and $\tau$ leptons.  Likewise, the decay of the atoms is hastened by the high probability of the constituent particles decaying.  
Nevertheless, if sufficient numbers of $\mu$ and $\tau$ pairs are produced in high energy astrophysical environments there may be significant  production of true muonium and true tauonium, despite the small probabilities.  This paper addresses this possibility.

We have calculated the pair production spectra of $\mu$ and $\tau$ leptons from photon-photon annihilation and electron-positron annihilation in astrophysical environments.  We have computed the cross-sections for radiative recombination and direct annihilation of the pairs, and the decay constants for the various allowable decays, and the wavelengths and energies of the recombination and annihilation signatures.  In this way we have
calculated the probabilities for the formation of true muonium and true tauonium, and the branching ratios for the various observable signatures.

We have estimated the expected fluxes from accretion discs around microquasars and active galactic nuclei, and from interactions of jets with clouds and stars.  We find that accretion discs around stellar mass black holes in our own Galaxy should have observable signatures at X-ray and $\gamma$-ray energies that are in principle observable with current observatories.

\end{abstract}

\maketitle

% This paper may be helpful as a compilation, a review of the likelihood of detection of various sources.
% But to really add something it would be good to apply the probabilities, even for unlikely reactions,
% to astrophysical events such as SNe and GRBs

\section{Introduction}
\label{sec:intro}

In the years 1928 -- 1931 Dirac introduced the concept of antimatter\cite{dir28, dir30,dir31}.  In 1932 Anderson showed that these theoretical concepts were real with the detection of the positron\cite{and33}.  Then in 1934 Mohorovi{\v c}i{\'c} published a paper on the possibility of observing positronium (Ps) from astrophysical sources via its recombination lines\cite{moh34}.  This prescient paper went largely unnoticed, and the existence of Ps was predicted independently and separately by Pirenne \cite{pir44}, Ruark\cite{rua45}, Landau (unpublished work referred to in \citet{ali45}) and Wheeler\cite{whe46}.

To date, the observational challenge laid down by Mohorovi{\v c}i{\'c}, viz.\ to detect astrophysical sources of Ps from its recombination lines, has not been met.  However the possibility of detection of Ps recombination lines has been raised on subsequent occasions inspired by advances in technology\cite{mcc84,burd97,ell09}.
Despite the non-detection of Ps recombination lines, there is clear evidence of its existence in the Galaxy from its annihilation signature\cite{lev78}; see the recent review by \citet{pra11}.

The observations of annihilating electrons and positrons in our Galaxy have raised several important questions for astrophysics, the foremost being what are the sources of the positrons?  Can they be explained with conventional astrophysics (e.g.\  high energy sources) or are more exotic sources required (e.g.\ annihilating dark matter)?  Why is the annihilation radiation centered so strongly on the nucleus of the Galaxy with relatively little disc emission?  How far do positrons disperse through the Galaxy before annihilation?

In this paper we address the question of what other observable signatures there are, arising from other leptonic atoms.  For example, it is possible in principle to form atoms of \mmu\ and \ttau.   However, in these cases the observable signatures are complicated by the fact that the constituent particles are themselves unstable, and may decay prior to annihilation or transitions\cite{avi79}.  It is important to note that even improbable signatures may be significant if the rate of production of the atom is high enough; this situation is often very relevant in astrophysical situations.  For example the hydrogen 21 cm emission line that results from  the spin-flip transition between the hyperfine splitting of the $^{2}$S ground state has been hugely exploited in radio astronomy since the 1950s\cite{ewe51,mull51}, although it has a transition probability of only  $2.9 \times 10^{-15}$ s$^{-1}$.

Even faint signatures would be of interest.  First, they would be the first such observations of \mmu\ or \ttau, either in astrophysics or in the laboratory.  Secondly they would place constraints on the energetics and rate of production of $\mu$ or $\tau$ in astrophysical sources.  Thirdly, the rate of \mmu\ or \ttau production, compared with estimates  of $\mu$ or $\tau$ production rates, would place constraints on the physical conditions during recombination.

Such measurements could be very useful.  Taking the example of Ps, the rate, distribution and energetics of formation are all well measured, but the identity of positron sources still remains unclear.  This is largely because it is difficult to reproduce the distribution of Ps emission, and also because there are many \e+\ candidates, which are difficult to distinguish observationally.  Observations, or even non-detection, of \mmu\ or \ttau\ could place further constraints on the competing models for the origin of Galactic positrons.   

The study of such processes is important, even if they are unlikely.  Anomalous spectral signatures are routinely discovered, for example the recent detection of an unidentified emission line in the X-ray spectra of clusters of galaxies at 3.5~keV\cite{boy14,bul14}.   We note that this line is not attributable to \mmu\ or \ttau, although these species are expected to produce X-ray emission lines, as we show.  Nevertheless, in such cases it is important to rule out prosaic, but poorly understood processes before accepting more exotic physics.  The formation of  \mmu\ and \ttau\ in astrophysical environments has so far received very little attention, but should not be rejected \emph{a priori} on the basis of the short lifetimes of the $\mu$ and $\tau$.

%If even faint signatures can be observed they will help to constrain fundamental areas of astrophysics and particle physics.  Detection would place constraints on the production mechanisms and energetics of the astrophysical sources of particle - anti-particle pairs;  e.g.\ different decay mechanisms are expected for different dark-matter annihilation models\cite{boe04}, and different \e+\ production mechanisms are expected for different astrophysical sources\cite{gue05}, only some of which will also produce heavier lepton pairs.  

The production of \mmu\ and \ttau\ in electron-positron colliders is difficult due to the rapid decay times of the particles, although methods have been proposed for the creation of \mmu\cite{brod09}.  It may be possible to bypass these difficulties if large numbers of leptonium atoms are produced in high energy astrophysical environments.  This is an important possibility since neither \mmu\ nor \ttau\ have ever been observed, and together they are the most compact pure QED systems that could exist.  An astrophysical detection would be more difficult to interpret than a collider detection, but would nevertheless constitute an important basic test of QED.

We wish to make a first step in addressing this challenge by calculating the expected observable signatures of astrophysical sources.  
Consider a source of particle - anti-particle lepton pairs.  Their possible fates are, (i) the particle immediately decays, if it is $\mu^{\pm}$ or $\tau^{\pm}$, (ii) it directly annihilates with its antiparticle, (iii) it radiatively recombines with a free particle, then annihilates, (iv) it forms a new atom via charge-exchange, then annihilates.  

Ps clearly has the highest probability of forming, since its constituent particles are stable, and can travel through the interstellar medium until they form Ps through charge-exchange or radiative recombination, although they may also directly annihilate in-flight or when thermalized.  The probabilities for Ps formation have been discussed in detail\cite{mcc84,gou89,wal96}, and Ps formation at the Galactic centre is confirmed by the observation of the 511 keV annihilation signature and the three-photon triplet annihilation continuum\cite{jea06}, from which a Ps formation fraction of $\approx 97$ per cent is inferred.  See \citet{pra11} for a recent review of Ps astrophysics. 

The formation of any atom having a $\mu$ or $\tau$ lepton is much less likely, since the decay times are so short that no travel of the particles is possible.  Thus, we only consider the formation of true muonium and true tauonium (hereafter referred to as M and T\footnote{The nomenclature for such systems is somewhat confusing.  A bound state of a particle and its anti-particle is called an onium, however the name muonium is already given to the combination of an anti-muon with an electron to form a hydrogen like atom, and by analogy tauonium refers to an anti-tauon and an electron.  Therefore the systems \mmu\ and \ttau\ are usually referred to as \emph{true} muonium and \emph{true} tauonium, however to avoid confusion, and for brevity, we prefer to use the symbols M and T}) \emph{in situ}, i.e.\ the particle - anti-particle pairs must be created with an energy less than the ionisation energy, so that the particle forms immediately.    A summary of the basic properties of M, T, and Ps, to which frequent comparisons are made, is given in Table~\ref{tab:summ}; the calculations or references for  these values can be found throughout the paper, in the sections given in the final column of the table.

\begin{table*}
\caption{Summary of the main properties of Ps, M and T.}
\label{tab:summ}
\begin{tabular}{llllll}
&& Ps & M & T & Section\\ \hline
\multicolumn{2}{l}{Rest mass/annihilation energy (MeV)} & 0.511& 105.66& 1406.6 &\\
\multicolumn{2}{l}{Ionisation energy (eV)} &6.8&1784.1&23751.4 & \S\ref{sec:mechanisms}\\
\multicolumn{2}{l}{Bohr radius (m)} &$1.058 \times 10^{-10}$ &$5.199 \times 10^{-13}$& $3.044 \times 10^{-14}$ & \S\ref{sec:decay_rad} \\
\multicolumn{2}{l}{Decay time of constituent particles (s)} &$\infty$&$2.197 \times 10^{-6}$ & $2.874 \times 10^{-13}$& \S\ref{sec:mutaudecay}\\
$\gamma\gamma$ Annihilation time (ground state) (s)& Singlet&$1.2 \times 10^{-10}$&$6.0 \times 10^{-13}$& $3.6 \times 10^{-14}$&\S\ref{sec:decay_ann}\\
& Triplet &$1.4 \times 10^{-7}$& $6.7 \times 10^{-10}$ &$4.0 \times 10^{-11}$ & \S\ref{sec:decay_ann}\\
e$^{-}$e$^{+}$ Annihilation time (ground state) (s)& Triplet &--&$1.8 \times 10^{-12}$& $1.1 \times 10^{-13}$& \S\ref{sec:decay_ann}\\
Recombination time ($nL \rightarrow n'L'$) (s) & $2\ 1 \rightarrow 1\ 0$ (Lyman $\alpha$)&$3.191\times 10^{-9}$ &
   $1.543\times 10^{-11}$ & $9.18\times 10^{-13}$ & \S\ref{sec:decay_rad} \\
&$3\ 1 \rightarrow 1\ 0$ (Lyman $\beta$) &  $1.195\times 10^{-8}$ &  $ 5.780\times 10^{-11}$ & $3.44\times 10^{-12}$& \S\ref{sec:decay_rad}  \\
& $3\ 1 \rightarrow 2\ 0$ & $8.905\times 10^{-8}$ &$   4.307\times 10^{-10}$ & $2.56\times 10^{-11}$ & \S\ref{sec:decay_rad} \\
&$3\ 0 \rightarrow 2\ 1$ & $3.166\times 10^{-7}$ &   $1.531\times 10^{-9} $&$ 9.11\times 10^{-11}$ & \S\ref{sec:decay_rad} \\
& $3\ 2 \rightarrow 2\ 1$ & $3.092\times 10^{-8}$ &$   1.495\times 10^{-10} $& $8.89\times 10^{-12}$ & \S\ref{sec:decay_rad} \\
& $4\ 1 \rightarrow 2\ 0$ &$2.068\times 10^{-7} $&$   9.999\times 10^{-10} $& $5.95\times 10^{-11}$& \S\ref{sec:decay_rad}  \\
& $4\ 0 \rightarrow 2\ 1$ &$7.753\times 10^{-7} $&$   3.750\times 10^{-9} $& $2.23\times 10^{-10} $& \S\ref{sec:decay_rad} \\
& $4\ 2 \rightarrow 2\ 1$ &$9.692\times 10^{-8}$ &$   4.687\times 10^{-10} $&$ 2.79\times 10^{-11}$& \S\ref{sec:decay_rad} \\
Recombination energies (keV) & Lyman $\alpha$ &$5.102 \times 10^{-3}$&1.055&17.7 & \S\ref{sec:decay_rad} \\
&Lyman $\beta$ &$6.047 \times 10^{-3}$&1.250& 21.0& \S\ref{sec:decay_rad} \\
& Balmer $\alpha$ &$9.448 \times 10^{-4}$&0.195&3.3& \S\ref{sec:decay_rad} \\
&Balmer $\beta$ &$1.276 \times 10^{-3}$&0.264& 4.4& \S\ref{sec:decay_rad} 
\end{tabular}
\end{table*}

To calculate the likelihood of forming M and T, we begin by reviewing the possible mechanisms for pair-production which are relevant in astrophysical environments (section~\ref{sec:mechanisms}).  We then use the relevant cross-sections for pair-production, to calculate the pair-production spectra, and from these calculate the fraction of pairs which have sufficiently low energy to form an onium immediately, i.e.\ the fraction of pairs produced with kinetic energy less than the ionisation energy of the atom (section~\ref{sec:pairprodspec}).  These pairs may either directly annihilate, or radiatively recombine, and we calculate the branching fraction for these processes in section~\ref{sec:gould}.  For those that do radiatively recombine, we calculate the cross-sections for radiative recombination, and hence the fractions which recombine into the $nL$th quantum level in section~\ref{sec:wallyn}.  Then in section~\ref{sec:decay} we calculate the different decay channels and in section~\ref{sec:branch} the resulting branching ratios for an onium in the $n$th level, and thus the total branching fractions for specific observational signatures.  The signatures themselves are discussed in section~\ref{sec:signatures}.  This brings us to a position where, for a given rate of pair-production, we can estimate the luminosity of each of the different decay channels, and we apply these calculations to specific astrophysical sources in section~\ref{sec:astrophys}.  We give our conclusions in section~\ref{sec:discuss}.

%The first step in addressing this challenge is to calculate  the lifetimes and reaction rates for various leptonic atoms, annihilations, decays and transitions, which we do in section~\ref{sec:theory}.  Following this we calculate the observable signatures of these reactions in section~\ref{sec:signatures}.  We then consider examples of astrophysical sources and estimate the strengths of the different observable signatures to be expected based in section~\ref{sec:sources}.  Finally we discuss our results and the prospects for detection in section~\ref{sec:discuss}.

\section{Pair-production processes in astrophysical environments}
\label{sec:mechanisms}

The formation of leptonium other than Ps is not complicated by possibilities of escape, in-flight formation, charge-exchange etc.; the lifetimes of the particles themselves are so short (\S~\ref{sec:decay}), that if an atom is not formed immediately after the creation of a particle, then the particle itself will decay before any of these processes can occur.  Ultra-relativistic particles may live longer due to time dilation, but as for \e+\cite{bus79}, the annihilation signature will give rise to  very weak continuum emission.

We consider the following production processes for  leptons, $l$,
\begin{eqnarray}
\pi^{-} \rightarrow l^{-} + \overline{\nu}_{l}, \label{eqn:piminusmu}\\
\pi^{+} \rightarrow l^{+} + \nu_{l} \label{eqn:piplusmu}, \\
\gamma + \gamma \rightarrow l^{-} + l^{+}, \label{eqn:mupair}\\
e^{-} + e^{+} \rightarrow l^{-} + l^{+} \label{eqn:mueepair}.
\end{eqnarray}
The pion decay processes will occur in collisions of highly energetic ($>200$ MeV) cosmic rays with protons in the ISM\cite{gue05}.  The third process requires very high photon energies and will occur in the high energy environments surrounding compact objects, such as black-holes and neutron stars\cite{gue05}.  The last process may occur if the \e+\ have large kinetic energy, and collide with e$^{-}$ in the ISM, e.g.\ if an e$^{\pm}$ jet from an AGN or microquasar collides with an interstellar cloud of gas or a star. 

We can immediately dismiss the pion decay processes as irrelevant for $\tau$ leptons, since the $\tau$ are more massive than the $\pi^{\pm}$.  We can also dismiss the pion decay processes as irrelevant for the formation of M, on two grounds.  First, only a single particle is produced, and the probability of radiative recombination with another free $\mu$ before it decays is negligible.  Secondly, the energetics are unfavorable.  Since the reaction involves a single particle of known mass decaying into two particles, the energetics are precisely determined in the zero-momentum frame; the $\mu$ is created with a kinetic energy of 4.12~MeV, which is much higher than the ionisation energy of M, $\approx 1.4$~keV.  Thus the created $\mu$ must first lose energy, then collide with its antiparticle, before either particle decays, the probability of which is negligible.  

This leaves us with photon-photon pair production and electron-positron pair production. In both these processes the $\mu$ (or $\tau$) will be created with typical energies much higher than the ionisation energy of M (or T) ($\approx 1.4$~keV for M, and $\approx 23.7$~keV for T).  Only the small fraction of pairs whose total kinetic energy in the zero momentum (z.m.) frame is less than the ionisation energy can form an onium,  i.e.,
\begin{eqnarray}
T_{1}+T_{2} \le E_{\rm ion}, \label{eqn:oniumenergy} \\
2 (\gamma-1) m c^{2} \le \frac{\mu q^{4}}{32 \pi^{2} \epsilon_{0}^{2} \hbar^{2}} ,
\end{eqnarray}
where the right hand side is the ionisation energy in S.I. units, $m$ is the mass of the lepton, $\mu$ is the reduced mass of the atom, $q$ is the electron charge, and $\gamma$ is the Lorentz factor of the produced pairs in the z.m.\ frame.  Therefore the maximum Lorentz factor in the z.m.\ frame that the particles can have and still form an onium is,
\begin{equation}
\label{eqn:gammalimit}
\gamma_{\rm lim}=1+\frac{q^{4}}{128 \pi^{2} c^{2} \epsilon_{0}^{2} \hbar^{2}}=1+6.656 \times 10^{-6}.
\end{equation}
Thus, only very low energy pairs can produce an onium, and $\gamma_{\rm lim}$ is independent of mass.  We now estimate the fraction of pairs produced with $\gamma < \gamma_{\rm lim}$, for various formation mechanisms.

\section{The pair-production spectrum}
\label{sec:pairprodspec}

Throughout this section we will refer to $\mu$ and M, but the arguments are identical for $\tau$ and T.

\subsection{Pair production by photon-photon annihilation}
\label{sec:gg}

For two photons with energy $E_{1}$ and $E_{2}$ colliding at an angle $\theta$, pair production will occur if,
\begin{equation}
\label{eqn:pairprod}
E_{1}E_{2} \ge \frac{2 \left( m c^{2}\right)^{2}}{1-\cos \theta},
\end{equation}
where  $m=2\mu$ is the mass of an individual particle.  The total energy of either produced particle, $E_{0}$ is given by,
\begin{equation}
\label{eqn:pairprodenergy}
E_{0}^{2}=\gamma^{2} m^{2} c^{4} = \frac{E_{1}E_{2} (1 - \cos \theta)}{ 2},
\end{equation}
where $\gamma$ is the Lorentz factor of either of the created particles in the z.m.\ frame.  

Many authors\cite{gou67,bon71,aga83,cop90,boe97} have evaluated the pair-production spectrum (usually of e$^{\pm}$) based on various assumptions about the distributions of photon energies.  Here we repeat these calculations, except we integrate the resulting spectra in order to obtain the fraction of produced pairs which have sufficiently low energy to form M, i.e.\ those which obey equation~\ref{eqn:gammalimit}. 

The total cross-section for the reaction $\gamma + \gamma \rightarrow \mu^{-} + \mu^{+}$ is\cite{jau55},
\begin{eqnarray}
\sigma &= &F_{\rm c} \frac{q^{4}}{32 \pi \epsilon_{0}^{2} m^{2} c^{4}} \nonumber \\
&&(1-\beta^{2})\left((3-\beta^{4})\ln\frac{1+\beta}{1-\beta} - 2\beta(2-\beta^{2})\right) \label{eqn:ggsigma} 
\end{eqnarray}
\begin{equation}
F_{\rm c} =\frac{\pi\frac{\alpha}{\beta}}{1 - {\rm e}^{-\pi\frac{\alpha}{\beta}}}. \label{eqn:ssc}
\end{equation}
where $\beta=u/c$, and $u$ is the velocity of the outgoing particles in z.m.\ frame, and $F_{\rm c}$ is the Sommerfeld-Sakharov correction to the cross-section, due to the attractive Coulomb force between the created $\mu^{\pm}$, which applies near the threshold energy\cite{sak48,sak91,vol03}.
 Now consider a source of photons whose number density per unit energy is given by $n_{1}(E_{1})$.  Let these photons interact with an isotropic photon gas whose number density per unit energy is given by $n_{2}(E_{2})$.  The fraction of photons in the isotropic gas which lie within a differential cone at angle $\theta$ and width ${\rm d}\theta$ to the incoming photon is $\frac{1}{2} \sin \theta {\rm d}\theta$.
The reaction rate in the lab  frame, i.e.\ the number of pairs created per unit volume per unit time is thus given by,
\begin{eqnarray}
R_{\gamma \gamma} &= & 
 \int_{0}^{\infty} \int_{-1}^{1} \int_{E_{1 {\rm min}}}^{E_{1 {\rm max}}} \nonumber \\ &&n_{1}(E_{1}) n_{2}(E_{2}) \sigma(\beta) \frac{c}{2}(1 - \cos \theta)  \nonumber \\ 
&&{ {\rm d}E_{1}\ \rm d}\cos\theta\  {\rm d}E_{2} \label{eqn:rgg}
 \end{eqnarray}
 where the factor $c(1-\cos \theta)$ is the relative velocity of the incoming photons, and the lower limit on the integral over $E_{1}$, 
 \begin{equation}
E_{1 {\rm min}} =\frac{2 m^{2} c^{4}}{E_{2}(1 - \cos \theta)}
\end{equation}
 from equation~\ref{eqn:pairprod}, ensures pair production is possible.  The upper limit on $E_{1}$ is either 
 \begin{equation}
E_{1 {\rm max}} =E_{1 {\rm M}}=\frac{2 \gamma_{\lim}^{2} m^{2} c^{4}}{E_{2}(1 - \cos \theta)}
\end{equation}
to specify only the produced pairs which satisfy equation~\ref{eqn:gammalimit}, or 
 \begin{equation} 
E_{1 {\rm max}} =E_{1 \infty}=\infty
\end{equation}
for all produced pairs.
The fraction of pairs which can produce M, $f_{\gamma \gamma}$, is thus given by the ratio of R integrated over both these limits.

We  have evaluated equation~\ref{eqn:rgg} numerically for different cases of photon energy distribution.  Because we are interested in the ratio of $R$ for different limits in $\gamma$, and because $E_{1}$ and $E_{2}$ are related to $\gamma$ via equation~\ref{eqn:pairprodenergy}, the form for the second distribution $n(E_{2})$ does not matter; integrating over $E_{2}$ leads to a different constant which cancels  when taking the ratio.  Thus, $f_{\gamma \gamma}$ is identical for a specific power-law interacting with any isotropic photon gas.    The fraction $f_{\gamma \gamma}$ for different power-law indices are given in Table~\ref{tab:ggpl}.

\begin{table}
\center
\caption{The fraction of pairs produced via photon-photon annihilation with sufficiently low energy to form an onium, for photon distributions with a power-law energy density with index $-\alpha$.}
\label{tab:ggpl}
\begin{tabular}{lc}
$\alpha$ & $f_{\gamma \gamma}$\footnote{Since the fraction $f_{\gamma \gamma}$ is independent of mass this is the fraction for any onium, Ps, M or T.} \\ \hline
1& $2.0 \times 10^{-7}$\\
1.5 & $3.8 \times 10^{-7}$\\
2 & $6.1 \times 10^{-7}$\\
2.5& $8.9 \times 10^{-7}$
\end{tabular}
\end{table}

\subsection{Pair production by electron-positron annihilation}  
\label{sec:ee}

The total cross-section for the process $e^{-} + e^{+} \rightarrow \mu^{-} + \mu^{+}$ is given by\cite{per92},
\begin{equation}
\sigma =F_{\rm c} \frac{\pi  \alpha^{2} \hbar^{2} \beta  \left(\beta^{4}-4 \beta^{2}+3\right)}{6 M^{2} c^{2}} 
\end{equation}
%\begin{eqnarray}
%\sigma &=&  \frac{e^{4} \sqrt{\gamma^{2}-1}}{192  \epsilon_{0}^{2} c^{4} M^{3} \pi
%\gamma^{6} \sqrt{  M^2 \gamma^{2} -m^2}} \times \nonumber \\
%&&   \biggl(m^{2} (7 - 4 \gamma^{2}) +    4 M^{2} \gamma^{2} (\gamma^{2}-1)\biggr) \nonumber \\ \label{eqn:emucross}
%\end{eqnarray}
%
where $M$ is the muon mass, and $F_{\rm c}$ is the Sommerfeld-Sakharov factor given by equation~\ref{eqn:ssc}.
The fraction of those muons produced which have sufficiently low energy that they could immediately form M now given by 
\begin{eqnarray}
R_{\rm ee} &= & 
 \int_{0}^{\infty} \int_{-1}^{1} \int_{E_{1 {\rm min}}}^{E_{1 {\rm max}}} \nonumber \\ &&n_{1}(E_{1}) n_{2}(E_{2}) \sigma(\beta)  \frac{v}{2} \nonumber \\ 
&&{ {\rm d}E_{1}\ \rm d}\cos\theta\  {\rm d}E_{2} \label{eqn:ree}
 \end{eqnarray}
 where $v$ is the relative speed of the electron and positron in the lab frame and $n_{1}(E_{1})$ and $n_{2}(E_{2})$ are now the distributions of the lab frame number density per unit energy of the electrons and positrons.
 
 The lab frame energies $E_{1}$ and $E_{2}$ are related to the Lorentz factors of the produced pairs by equation~\ref{eqn:eemmen}, and the relative velocity in the lab frame, $v$, is given by equation~\ref{eqn:eemmv}.  Therefore changing  variables to integrate over $\gamma$ instead of $E_{1}$, we can calculate $R_{{\rm ee}_{\gamma_{\rm lim}}}$ in the limit $\gamma_{\rm max} = \gamma_{\rm lim}$, and  $R_{{\rm ee}_{\infty}}$ in the limit $\gamma_{\rm max} = \infty$, and thus the fraction of produced pairs with sufficiently low energy to form M is given by,
 \begin{equation}
 \label{eqn:fMee}
 f_{\rm ee} = \frac{R_{{\rm ee}_{\gamma_{\rm lim}}}}{R_{{\rm ee}_{\infty}}}.
 \end{equation}
 %
 %The integrations were performed numerically using {\sc Mathematica}.  
 As for photon-photon annihilation the exact energy density of $E_{2}$ does not matter, since it cancels when taking the ratio, and the result is independent of the mass of the produced particles.  We have calculated  $f_{\rm ee}$ for various power-law indices, where we evaluated the integrals numerically, and we show the results in Table~\ref{tab:eemm}.
 
 \begin{table}
\center
\caption{The fraction of pairs produced via electron-positron annihilation with sufficiently low energy to form an onium, for electron distributions with a power-law energy density with index $-\alpha$.}
\label{tab:eemm}
\begin{tabular}{lc}
$\alpha$ & $f_{\rm ee}$\footnote{Since the fraction $f_{\rm ee}$ is independent of mass this is the fraction for any onium, Ps, M or T.}  \\ \hline
1& $5.7 \times 10^{-7}$\\
1.5 & $9.1 \times 10^{-7}$\\
2 & $1.3 \times 10^{-6}$ \\
2.5& $1.7 \times 10^{-6} $
\end{tabular}
\end{table}

\section{Cross-sections for radiative recombination and direct annihilation}
\label{sec:gould}

In the previous section we calculated the fraction of pairs produced, either via photon-photon annihilation, or by electron positron annihilation, which have sufficiently low energy to form Ps, M or T immediately. 
However, the fraction of these pairs which will  form an onium is  less still.  As for Ps\cite{gou89} there is a still a probability that some of these pairs will directly annihilate; though in the case of $\mu$ and $\tau$ we can neglect charge exchange since the probability of the particles decaying before meeting a neutral atom is large.

\begin{figure}
\centering \includegraphics[scale=0.6]{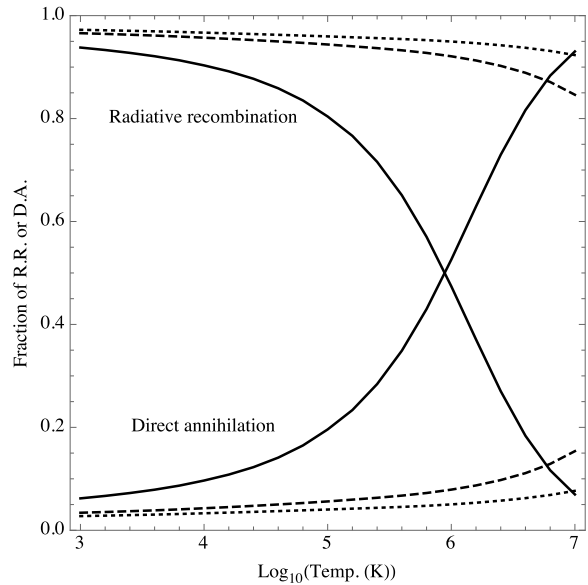}
\caption{The fraction of thermalised pairs which will either radiatively recombine (R.R.) or directly annihilate (D.A.), neglecting charge-exchange, as a function of temperature, for e$^{\pm}$ (continuous line), $\mu^{\pm}$ (dashed line), and $\tau^{\pm}$ (dotted line).}
\label{fig:rrda}
\end{figure}

We have calculated the total reaction rates for radiative recombination and direct annihilation following \citet{gou89} (see also \citet{gou72}).  Figure~\ref{fig:rrda} shows the fraction of pairs which radiatively recombine, $g_{\rm rr}$, or directly annihilate, $g_{\rm da}$, as a function of temperature.  For e$^{\pm}$ radiative recombination dominates below $\approx 10^{6}$ K, and for $\mu$ and $\tau$, radiative recombination dominates at all astrophysically relevant temperatures.  

\subsection{Cross-sections for radiative recombination onto the $nL$th level}
\label{sec:wallyn}

\begin{figure}
  \subfigure[The total cross section for levels  $nL$]{
\centering \includegraphics[scale=0.58]{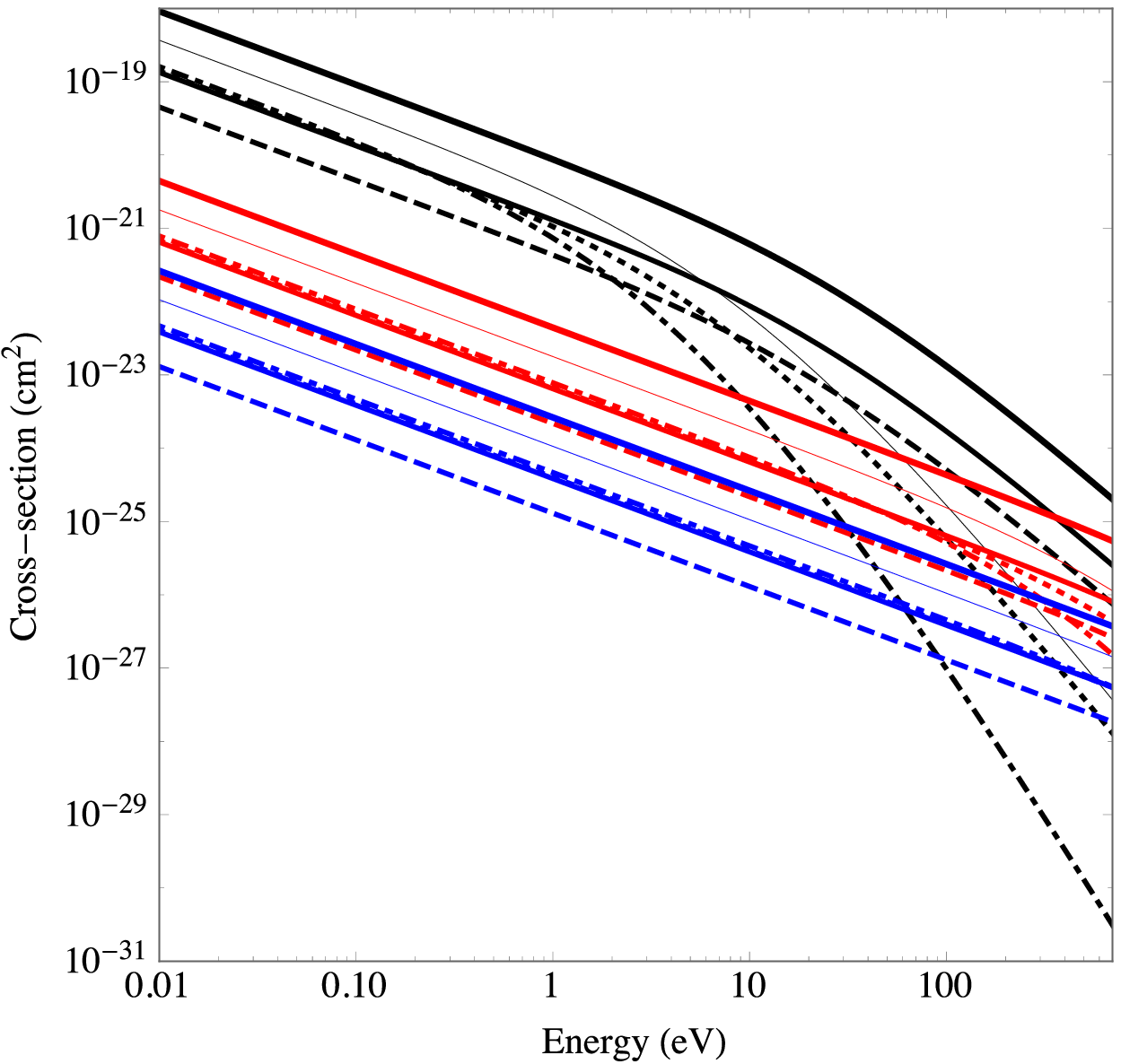}
}
  \subfigure[The total cross section for level $n$, summed over all the $L$ sub-levels.]{
\centering \includegraphics[scale=0.58]{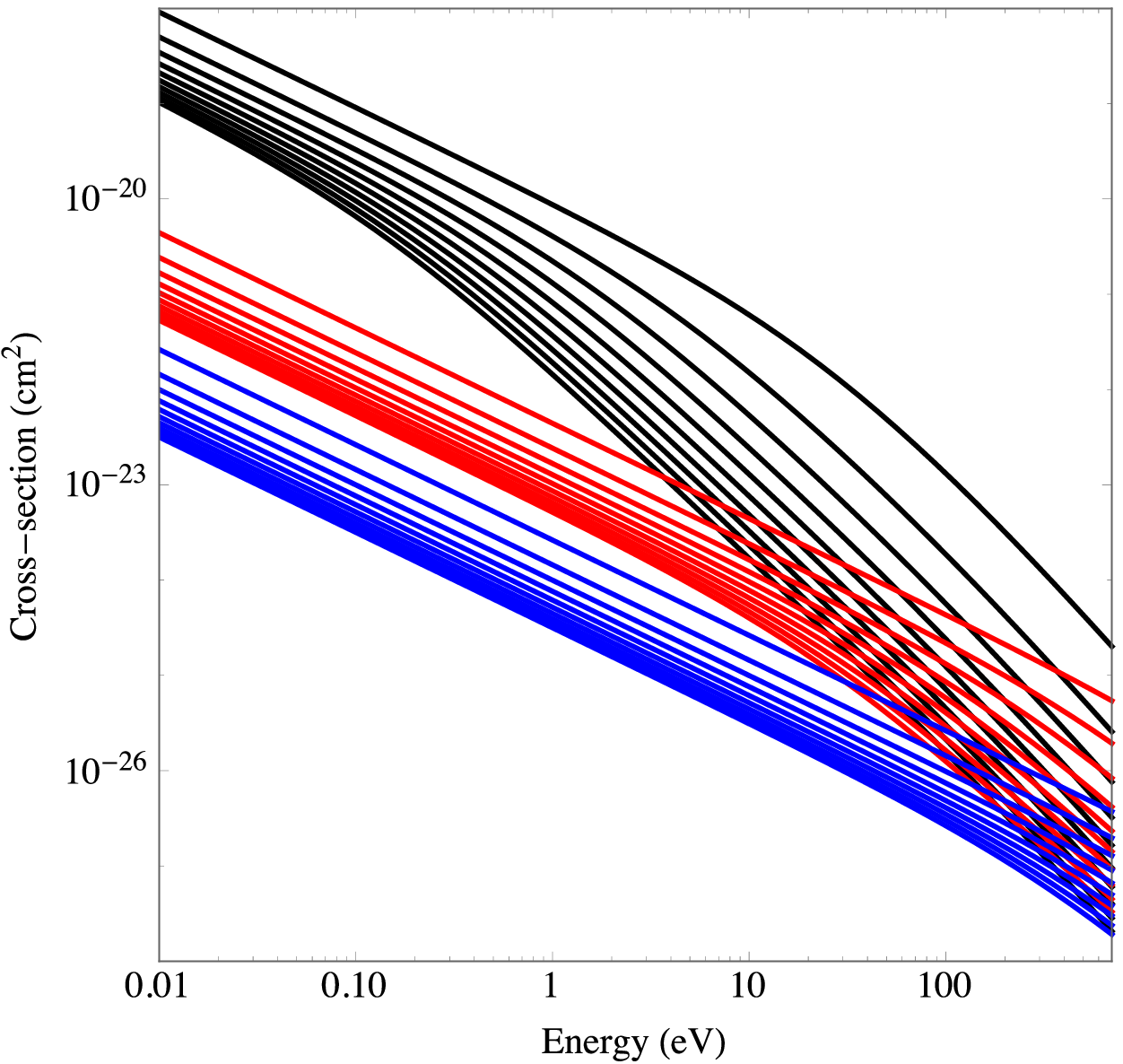}
}
\caption{(a ) The cross section for the radiative recombination of Ps (black), M (red) and T (blue), onto the quantum levels $nL$ for $n=1$ to 3 as a function of the relative energy of the combining particles.  The lines are coded (1,0) thick, (2,0) normal, (2,1) thin, (3,0) dashed, (3,1) dotted, (3,2) dot-dashed.  (b) The total cross sections for level $n$,  from  $n=1$ to 10, summed over all the $L$ sub-levels.}
\label{fig:sigman}
\end{figure}

For those pairs which do radiatively recombine to form an onium, we require the fraction which recombine into the $nL$th level, such that we may then determine the subsequent decay.
\citet{wal96} give the cross-sections for the formation of Ps via radiative recombination as a function of the quantum levels $n$ and $L$ and the relative energy of of the electron and positron.  We have repeated these calculations for Ps, M and T, and show the results in Figure~\ref{fig:sigman}.  Thus we may calculate the fraction of oniums which form in the $nL$th energy level via radiative recombination as 
\begin{equation}
\label{eqn:fsigmanL}
f_{\sigma_{nL}} = \frac{\sigma(nL)}{\sum\limits_{n=1}^{n=\infty} \sum\limits_{L=0}^{L=n-1} \sigma(nL)}.
\end{equation}
  In practice we cannot sum to $n=\infty$, and instead choose a suitably high value of $n=n_{\rm lim}$, such that the fraction $f_{\sigma_{nL}}$ asymptotes to a constant value.  
  %Figure~\ref{fig:sigmanfrac} shows $f_{\sigma_{n}}$ as a function of $n_{ \rm lim}$ for different relative energies and for $n=1,2,3$ for Ps, M and T.   
 We computed $f_{\sigma_{nL}}$ for various limits, $n_{\rm lim}$, and found that $n_{\rm lim} = 300$, 500 or 1000  is sufficient for Ps, M, or T respectively.  Using these limits we have calculated the fractions $f_{\sigma_{nL}}$ for Ps, M and T as a function of temperature, and give the results for the first six energy levels in Table~\ref{tab:sigmanfract}.
  
\begin{table*}
\caption{The fraction of those atoms which radiatively recombine to form Ps, M or T, which do so in the level $nL$.}
\label{tab:sigmanfract}
\begin{tabular}{ll|lllll|lllll|lllll}
&&\multicolumn{15}{c}{Temperature (K)} \\
&&\multicolumn{5}{c}{Ps}&\multicolumn{5}{c}{M}&\multicolumn{5}{c}{T} \\ 
 n & L & $10^{3}$& $10^{4}$& $10^{5}$& $10^{6}$& $10^{7}$ & $10^{3}$& $10^{4}$& $10^{5}$& $10^{6}$& $10^{7}$ & $10^{3}$& $10^{4}$& $10^{5}$& $10^{6}$& $10^{7}$ \\ \hline

   1 & 0 & 0.296 & 0.459 & 0.702 & 0.814 & 0.830 & 0.153 & 0.193
   & 0.263 & 0.396 & 0.628 & 0.123 & 0.145 & 0.182 & 0.244 &
   0.358 \\
 2 & 0 & 0.043 & 0.068 & 0.100 & 0.105 & 0.104 & 0.022 & 0.028
   & 0.039 & 0.059 & 0.093 & 0.018 & 0.021 & 0.027 & 0.036 &
   0.053 \\
 2 & 1 & 0.116 & 0.143 & 0.068 & 0.009 & 0.001 & 0.062 & 0.078
   & 0.105 & 0.140 & 0.102 & 0.050 & 0.058 & 0.073 & 0.097 &
   0.133 \\
 3 & 0 & 0.015 & 0.023 & 0.031 & 0.031 & 0.031 & 0.008 & 0.010
   & 0.013 & 0.020 & 0.029 & 0.006 & 0.007 & 0.009 & 0.012 &
   0.018 \\
 3 & 1 & 0.044 & 0.055 & 0.025 & 0.003 & 0.000 & 0.023 & 0.030
   & 0.040 & 0.053 & 0.038 & 0.019 & 0.022 & 0.028 & 0.037 &
   0.051 \\
 3 & 2 & 0.048 & 0.037 & 0.003 & 0.000 & 0.000 &
   0.027 & 0.034 & 0.045 & 0.046 & 0.010 & 0.022 & 0.026 &
   0.032 & 0.042 & 0.049 \\
 4 & 0 & 0.007 & 0.010 & 0.013 & 0.013 & 0.013 & 0.004 & 0.004
   & 0.006 & 0.009 & 0.013 & 0.003 & 0.003 & 0.004 & 0.006 &
   0.008 \\
 4 & 1 & 0.021 & 0.026 & 0.011 & 0.001 & 0.000 & 0.011 & 0.014
   & 0.019 & 0.025 & 0.017 & 0.009 & 0.011 & 0.013 & 0.018 &
   0.024 \\
 4 & 2 & 0.030 & 0.023 & 0.002 & 0.000 & 0.000 &
   0.017 & 0.021 & 0.028 & 0.029 & 0.006 & 0.014 & 0.016 &
   0.020 & 0.026 & 0.030 \\
 4 & 3 & 0.020 & 0.007 & 0.000 & 0.000 &
   0.000 & 0.013 & 0.016 & 0.020 & 0.013 &
   0.001 & 0.010 & 0.012 & 0.015 & 0.019 & 0.017 \\
 5 & 0 & 0.004 & 0.006 & 0.007 & 0.007 & 0.007 & 0.002 & 0.003
   & 0.003 & 0.005 & 0.007 & 0.002 & 0.002 & 0.002 & 0.003 &
   0.005 \\
 5 & 1 & 0.012 & 0.014 & 0.006 & 0.001 & 0.000 & 0.006 & 0.008
   & 0.011 & 0.014 & 0.009 & 0.005 & 0.006 & 0.007 & 0.010 &
   0.013 \\
 5 & 2 & 0.018 & 0.014 & 0.001 & 0.000 & 0.000 &
   0.010 & 0.013 & 0.017 & 0.017 & 0.004 & 0.008 & 0.010 &
   0.012 & 0.016 & 0.019 \\
 5 & 3 & 0.017 & 0.007 & 0.000 & 0.000 &
   0.000 & 0.011 & 0.014 & 0.017 & 0.012 &
   0.001 & 0.009 & 0.011 & 0.013 & 0.017 & 0.015 \\
 5 & 4 & 0.008 & 0.001 & 0.000 & 0.000 & 0.000 & 0.006 & 0.008 & 0.009 &
   0.003 & 0.000 & 0.005 & 0.006 & 0.007 & 0.009 & 0.005 \\
 6 & 0 & 0.002 & 0.003 & 0.004 & 0.004 & 0.004 & 0.001 & 0.002
   & 0.002 & 0.003 & 0.004 & 0.001 & 0.001 & 0.001 & 0.002 &
   0.003 \\
 6 & 1 & 0.007 & 0.008 & 0.003 & 0.000 & 0.000 & 0.004 & 0.005
   & 0.007 & 0.008 & 0.005 & 0.003 & 0.004 & 0.005 & 0.006 &
   0.008 \\
 6 & 2 & 0.012 & 0.009 & 0.001 & 0.000 & 0.000 &
   0.007 & 0.008 & 0.011 & 0.011 & 0.002 & 0.005 & 0.006 &
   0.008 & 0.010 & 0.012 \\
 6 & 3 & 0.013 & 0.005 & 0.000 & 0.000 &
   0.000 & 0.008 & 0.010 & 0.013 & 0.009 &
   0.000 & 0.007 & 0.008 & 0.010 & 0.012 & 0.011 \\
 6 & 4 & 0.009 & 0.001 & 0.000 & 0.000 &0.000 & 0.007 & 0.009 & 0.010 &
   0.004 & 0.000 & 0.006 & 0.007 & 0.008 & 0.010 & 0.006 \\
 6 & 5 & 0.003 & 0.000 & 0.000 & 0.000 & 0.000 & 0.003 & 0.004 & 0.004 &
   0.001 & 0.000 & 0.002 & 0.003 & 0.004 & 0.004
   & 0.001 \\

    \end{tabular}
 \end{table*}
 %\begin{figure*}
 %\subfigure[0.1 eV]{
 %\includegraphics[scale=0.4]{sigmanps01.eps}
 %\includegraphics[scale=0.4]{sigmanM01.eps}
 %\includegraphics[scale=0.4]{sigmanT01.eps}
 %}
% \subfigure[1 eV]{
 %\includegraphics[scale=0.4]{sigmanps1.eps}
 %\includegraphics[scale=0.4]{sigmanM1.eps}
 %\includegraphics[scale=0.4]{sigmanT1.eps}
 %}
 %\subfigure[10 eV]{
 %\includegraphics[scale=0.4]{sigmanps10.eps}
 %\includegraphics[scale=0.4]{sigmanM10.eps}
 %\includegraphics[scale=0.4]{sigmanT10.eps}
 %}
 %\subfigure[100 eV]{
 %\includegraphics[scale=0.4]{sigmanps100.eps}
 %\includegraphics[scale=0.4]{sigmanM100.eps}
 %\includegraphics[scale=0.4]{sigmanT100.eps}
 %}
 %\caption{The fraction $\sigma(n)/\sum\limits_{n=1}^{n_{\rm lim}}$ as a function of $n_{\rm lim}$, for $n=1, 2 3$, represented by points of increasing size, respectively.  The columns are for Ps, M and T from left to right.  The rows are for different relative energies as labelled.}
% \label{fig:sigmanfrac}
 %\end{figure*}

\section{Decay channels}
\label{sec:decay}

We have now calculated the fraction of pairs produced via photon-photon (\S~\ref{sec:gg}), or electron-positron (\S~\ref{sec:ee}), annihilation which have sufficiently low energy to form Ps, M or T, and of those which do, the fraction which radiatively recombine or directly annihilate (\S~\ref{sec:gould}), and of those which radiatively recombine to form an onium, the fraction which form in the $nL$th level (\S~\ref{sec:wallyn}).

An onium in the $nL$th level may decay via annihilation of its constituent particles, if $L=0$; it may radiatively transition to a lower quantum level if $n >1$; or if it is M or T, either one of the individual constituent particles may decay from any level.  We  wish to calculate the branching fractions for these various decay channels, and we begin by calculating the lifetimes for each process.    

\subsection{Annihilation}
\label{sec:decay_ann}

Invariance under charge-conjugation leads to the following selection rule for the decay of an onium into $n$ photons,
\begin{equation}
(-1)^{l+s} = (-1)^{n},
\end{equation} 
where $l$ and $s$ are the orbital angular momentum and spin quantum numbers respectively\cite{yan50,wol52}.   Furthermore, since the electron and positron (or $\mu^{-}$ and $\mu^{+}$ etc.) only overlap in the $L=0$ state, annihilation into photons is only possible (barring negligible higher order decay processes) from the singlet $^{1}S_{0}$ state or from the triplet $^{3}S_{1}$ state.  The $^{1}S_{0}$ will decay into an even number of photons, with 2 being the most probable. The $^{3}S_{1}$ state will decay into an odd number of photons, with 3 being the most probable, since 1 is forbidden due to conservation of momentum.  However, for M and T the triplet state can also decay via one photon, which then decays into e$^{\pm}$ pairs, i.e.\ M~$\rightarrow \gamma^{*} \rightarrow {\rm   e}^{+}{\rm e}^{-}$.

The decay rate of an onium in the $n$th level of the singlet state is given by\cite{dir30b},
\begin{equation}
\Gamma_{1} = \frac{1}{n^{3}} \frac{\alpha^{5} \mu c^{2}}{\hbar},
\end{equation}
where $\mu$ is the reduced mass of the onium.

For triplet states the lowest order decay rate is \cite{ore49,berk80},
\begin{equation}
\Gamma_{3}=\frac{1}{n^{3}}\frac{2}{9\pi}(\pi^{2}-9)\left(\frac{2 \mu c^{2}}{\hbar}\right)\alpha^{6}.
\end{equation}
For M and T the decay of the triplet state into electron-positron pairs has a decay rate given by\cite{brod09},
\begin{equation}
\Gamma_{e^{+}e^{-}} = \frac{\alpha^{5} \mu c^{2}}{3 \hbar n^{3}}.
\end{equation}

\subsection{Decay}
\label{sec:mutaudecay}

Both the $\mu$ and the $\tau$ lepton are intrinsically unstable and will decay to lighter particles.  
%We take the decay lifetimes to be, $T_{\mu} = 2.197 \times 10^{-6}$~s and $T_{\tau} = 2.874 \times 10^{-13}$~s \cite{per92}.
We take the decay rates to be, $T_{\mu} =1/\Gamma_{\mu} = 2.197 \times 10^{-6}$~s and $T_{\tau} = 1/\Gamma_{\tau}=2.874 \times 10^{-13}$~s \cite{per92}.

\subsection{Radiation}
\label{sec:decay_rad}

The coefficient for a radiative transition of a hydrogenic atom from the state $n' L'$ to a lower state $nL$, $A_{n'L',nL}$ is given by\cite{gre57,pen64,bro71}
 \begin{equation}
A_{n'L',nL} = 4 \pi^{2} \nu^{3} \left(\frac{ 8 \pi \alpha a_{0}^{2}}{3 c^{2}}\right) \frac{{\rm Max}(L,L')}{(2L'+1)} \left|p(n'L',nL)\right|^{2},
\end{equation} 
where $\nu$ is the frequency of the transition,
\begin{equation}
\nu = c R_{\infty} \left(\frac{1}{n^{2}} - \frac{1}{n'^{2}}\right),
\end{equation}
$R_{\infty}$ is the Rydberg constant for the particular onium,
\begin{equation}
R_{\infty}=\frac{(\mu e^{4})}{8 \epsilon_{0}^{2} c h^{3}}
\end{equation}
and $a_{0}$ is the Bohr radius of the onium,
\begin{equation}
a_{0}=\frac{4 \pi \epsilon_{0} \hbar^{2}}{\mu e^{2}}.
\end{equation}

The dipole matrix elements are given by\cite{gor29},
\begin{widetext}
\begin{eqnarray}
|p(n'L-1,nL|^{2} &=& \Biggl( \frac{(-1)^{n'-1}}{4(2L-1)!} \sqrt{\frac{(n+L)!(n'+L-1)!}{(n-L-1)!(n'-L)!}} 
\frac{(4nn')^{L+1}}{(n+n')^{n+n'}} (n-n')^{n+n'-2L-2} \Biggr. \nonumber \\
&&  \times \biggl( \ _{2}F_{1}\left[-n+L+1,-n'+L,2L,\frac{-4nn'}{(n-n')^{2}}\right] \biggr. \nonumber \\
&& -\left(\frac{n-n'}{n+n'}\right)^{2} \ _{2}F_{1} \left[-n+L-1,-n'+L,2L,\frac{-4 nn'}{(n-n')^{2}}\right] \biggl. \biggr) \Biggl.\Biggr)^{2},
\end{eqnarray}
\end{widetext}
which is correct when $L>L'$; if $L<L'$ then $nL$ and $n'L'$ are swapped.  These can be adapted for any onium simply by replacing the reduced mass, $\mu$\cite{wal96}.

The radiative coefficient for a transition from a level $n'L'$ to \emph{any} lower level is thus given by,
\begin{equation}
A_{n'L'} = \sum\limits_{n=1}^{n=n'-1} \sum\limits_{L=L' \pm 1} A_{n'L',nL}. 
\end{equation}

\section{Branching ratios}
\label{sec:branch}

\subsection{Branching ratios for a particular level, $nL$}

Consider leptonium in the level $nL$.  It may decay via annihilation (two photon, three photon or $e^{\pm}$, depending on the state and the type of atom), radiative transition, or the decay of either constituent particle (for M and T).  The decay rates for these channels were calculated in the previous section.  Thus we may now calculate the probabilities for each process.

Recalling that annihilation can only take place from the states with $L=0$, and also that the annihilation mechanisms differ for singlet states, which decay into two photons, and triplet states, which decay either into three photons, or into $e^{\pm}$ pairs via a single photon, the total decay constant for leptonium in a level $nL$ via \emph{any} decay process is given by,
\begin{eqnarray}
A_{nL_{\rm sing}} = &A_{nL} +\delta(L)\Gamma_{1} + 2 \Gamma_{\mu,\tau},& ^{1}S_{0}\label{eqn:AnL1}\\
A_{nL_{\rm trip}} = &A_{nL} +\delta(L)\Gamma_{3}+\delta(L)\Gamma_{ee} + 2 \Gamma_{\mu,\tau},& ^{3}S_{1},\label{eqn:AnL3}
\end{eqnarray}
 where $\delta(L)$ is a function which is 1 when $L=0$, and zero otherwise.

Therefore the probability of making a particular radiative transition from a level $n'L'$ to a lower level $nL$ is given by,
\begin{eqnarray}
P_{n'L',nL_{\rm sing}} &= &\frac{1}{4}\frac{A_{n'L',nL}}{A_{n'L'_{\rm sing}}} \\
P_{n'L',nL_{\rm trip}} &= & \frac{3}{4}\frac{A_{n'L',nL}}{A_{n'L'_{\rm trip}}}, \\
P_{n'L',nL_{\rm tot}} &= &P_{n'L',nL_{\rm sing}}  + P_{n'L',nL_{\rm trip}} 
\end{eqnarray}
where the leading factors of $1/4$ and $3/4$ are the probabilities that the onium will form in the singlet or triplet state, respectively.

Similarly the probability for leptonium in the level $n'L'$ to decay via two-photon annihilation is,
\begin{equation}
P_{n'L'_{\gamma\gamma}} = \frac{1}{4}\frac{\delta(L')\Gamma_{1}}{A_{n'L'_{\rm sing}}}.
\end{equation}
Likewise for three photon annihilation,
\begin{equation}
P_{n'L'_{\gamma\gamma\gamma}} = \frac{3}{4}\frac{\delta(L')\Gamma_{3}}{A_{n'L'_{\rm trip}}},
\end{equation}
and annihilation into an electron-positron pair
\begin{equation}
P_{n'L'_{\rm ee}} = \frac{3}{4}\frac{\delta(L')\Gamma_{ee}}{A_{n'L'_{\rm trip}}}.
\end{equation}
Finally the probability that either constituent particle will decay is
\begin{equation}
P_{n'L'_{\rm decay}} = 2\left(\frac{\Gamma_{\mu,\tau}}{4}\frac{1}{ A_{n'L'_{\rm sing}}} + \frac{3}{4}\frac{\Gamma_{\mu,\tau} }{A_{n'L'_{\rm trip}}}\right)
\end{equation}
where $\Gamma_{\mu,\tau}$ is the decay rate of the $\mu$ or $\tau$ as appropriate.

\subsection{Total branching ratios}

The previous subsection gave the branching ratios for the various decays from a particular state $nL$.  We now need to consider the probability of leptonium being in the state $nL$ in the first place.  There are two possibilities for populating the state: it may recombine directly into the state $nL$, or it may recombine into a higher state and cascade down into $nL$.  That is, we assume Case A recombination\cite{ost06}; because leptonium is short-lived, there is no resonant scattering of emission lines from Ps, M or T, through the excitation and re-emission of surrounding Ps, M or T atoms.  For the same reason, we also assume that collisional excitation of the atoms is negligible.  

\subsubsection{Cascade probabilities}

The probability of cascading from a level $n'L'$ to a lower level $nL$ via \emph{all} cascade paths, $C_{n'L',nL}$, is calculated using an iterative procedure\cite{pen64}, such that,
%\footnote{Both \citet{pen64} and \citet{wal96} have an error in the limits on the sum over $L''$.},
%
\begin{widetext}
\begin{equation}
C_{n'L',nL_{\rm sing,trip}} = 
\begin{cases}
1,& n=n' \wedge L=L' \\ 
\sum\limits_{n''=n}^{n'-1} \sum\limits_{L''=L' \pm 1}  P_{n'L',n''L''_{\rm sing,trip}} C_{n''L'',nL_{\rm sing,trip}},& n\neq n' \vee L\neq L' , 
\end{cases}
\end{equation}
\end{widetext}
where the subscripts `sing' and `trip' are used to specify the calculations for the singlet state or triplet state, an unfortunate but necessary complication since the decay paths are different for each.

\subsubsection{Population of the state, $nL$}

Meanwhile the state is depopulated by the processes described in section~\ref{sec:decay}.  Therefore, in equilibrium the population of any state, $N_{nL}$ is given by\cite{sea59,pen64},
\begin{eqnarray}
\sum\limits_{n'=n}^{\infty}\sum\limits_{L'=0}^{n'-1} N_{\rm pairs} f_{\rm ion} g_{\rm rr} f_{\sigma_{nL}} C_{n'L',nL_{\rm sing,trip}} \nonumber \\
= N_{nL_{\rm sing,trip}} A_{nL_{\rm sing,trip}},
\end{eqnarray}
where $N_{\rm pairs}$ is the rate at which pairs are produced, 
$f_{\rm ion}$ is the fraction of pairs produced with energy less than the ionization energy, i.e.\ $f_{\gamma \gamma}$ or $f_{ee}$ in sections~\ref{sec:gg} and \ref{sec:ee} above, 
$g_{\rm rr}$ is the fraction of those pairs which radiatively recombine (\S~\ref{sec:gould}, Figure~\ref{fig:rrda}), 
 $f_{\sigma_{nL}}$ is the fraction of those pairs which form in level $nL$, (\S~\ref{sec:wallyn}, equation~\ref{eqn:fsigmanL}),
 and again the equation has two forms reflecting the different decay paths of the singlet and triplet states.  

Let us then define $f_{nL}$ as the relative population of the level $nL$,  i.e.,
\begin{eqnarray}
f_{nL_{\rm sing,trip}}&=&\frac{N_{nL_{\rm sing,trip}}}{N_{\rm pairs} f_{\rm ion} g_{\rm rr} }\nonumber \\
& =& \frac{\sum\limits_{n'=n}^{\infty} \sum\limits_{L'=0}^{n'-1}f_{\sigma_{n'L'}}  C_{n'L',nL_{\rm sing,trip}} }{A_{nL_{\rm sing,trip}}}, \label{eqn:fnL}
\end{eqnarray}
which is the fraction of leptonium in the level $nL$ for a population whose formation and decay rates are in equilibrium.

\subsubsection{Branching ratios}

The final step in calculating the total branching ratios is simply to multiply the fractional population each state, $f_{nL}$, by the appropriate decay constant, and sum over all levels.  Thus,
\begin{widetext}
\begin{align}
P_{\rm Ly\alpha}  &=&& A_{21,10}\left(\frac{1}{4}  f_{21_{\rm sing}} + \frac{3}{4}  f_{21_{\rm trip}}\right), & {\rm Lyman}\ \alpha \label{eqn:pla}\\
P_{\rm Ly\beta} &= && A_{31,10}\left(\frac{1}{4}  f_{31_{\rm sing}} + \frac{3}{4}  f_{31_{\rm trip}}\right), &{\rm Lyman}\ \beta \\
P_{\rm Ba\alpha}  &=&&  A_{31,20}\left(\frac{1}{4}  f_{31_{\rm sing}} + \frac{3}{4}  f_{31_{\rm trip}}\right) + A_{30,21}\left(\frac{1}{4}  f_{30_{\rm sing}} + \frac{3}{4}  f_{30_{\rm trip}}\right), &{\rm Balmer}\ \alpha \\
P_{\rm Ba\beta}  &=&&  A_{41,20}\left(\frac{1}{4}  f_{41_{\rm sing}} + \frac{3}{4}  f_{41_{\rm trip}}\right) + A_{40,21}\left(\frac{1}{4}  f_{40_{\rm sing}} + \frac{3}{4}  f_{40_{\rm trip}}\right), & {\rm Balmer}\ \beta \\
P_{\gamma\gamma} &=&&\sum\limits_{n=0}^{\infty} \frac{1}{4}f_{n0_{\rm sing}} \Gamma_{1}(n), &{\rm Two\ photon\ annihilation} \label{eqn:pgamgam}\\
P_{\gamma\gamma\gamma} &=&&\sum\limits_{n=0}^{\infty} \frac{3}{4}f_{n0_{\rm trip}} \Gamma_{3}(n), &{\rm Three\ photon\ annihilation} \\
P_{\rm ee} &=&&\sum\limits_{n=0}^{\infty} \frac{3}{4}f_{n0_{\rm trip}} \Gamma_{ee}(n), &{\rm Electron-positron\ annihilation} \label{eqn:pelpos}\\
P_{\rm decay} &=&& \sum\limits_{n=0}^{\infty}\sum\limits_{L=0}^{n-1} 2 \Gamma_{\mu,\tau} \left(\frac{1}{4} f_{nL_{\rm sing}} + \frac{3}{4} f_{nL_{\rm trip}}\right),& {\rm Decay\ of\ either\ particle} \label{eqn:pdec}
\end{align}
\end{widetext}

All of the equations~\ref{eqn:pla} -- \ref{eqn:pdec}, contain either one or two infinite sums.  All include the infinite sum in equation~\ref{eqn:fnL} which accounts for recombination into any level up $n=\infty$ followed by cascade to the relevant level.  Equations~\ref{eqn:pgamgam} -- \ref{eqn:pdec} also include the sum over all levels up to $n=\infty$ which all contribute to the relevant decay process.  In practice we cannot compute these infinite sums, and instead compute equations~\ref{eqn:pla} -- \ref{eqn:pdec} up to certain limits on each sum.  We then alter these limits, and fit to the resulting curve or surface, and extrapolate our results to $n=\infty$.  The limits are $n=11$ -- $15$, for the capture-cascade sum, and $n=6$ -- $10$ for the sum over the decays.  

\begin{figure}
\centering \includegraphics[scale=0.58]{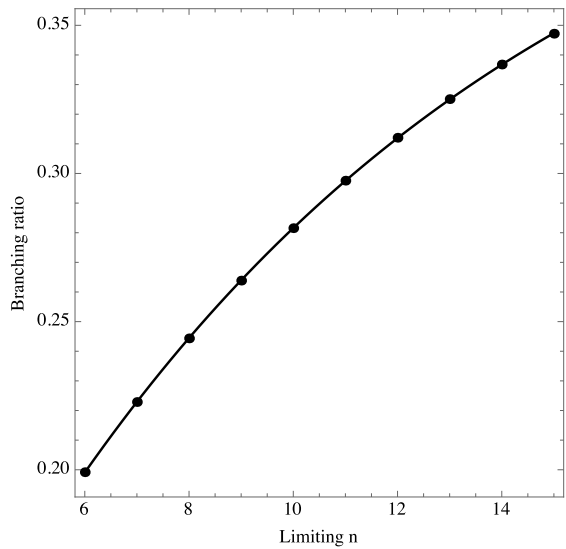}
\caption{The branching ratio of Ly $\alpha$ emission from M at a gas temperature of $T=10^{4}$~K as a function of the limiting value of the sum over $n$ in equation~\ref{eqn:fnL}, shown by the points.  We find good fits with a  function of the form given in equation~\ref{eqn:ffit1}, shown by the curve; in this case giving a limiting value as $n \rightarrow \infty$ of 0.45.}
\label{fig:MLyafit}
\end{figure}

\begin{figure}
\centering \includegraphics[scale=0.31]{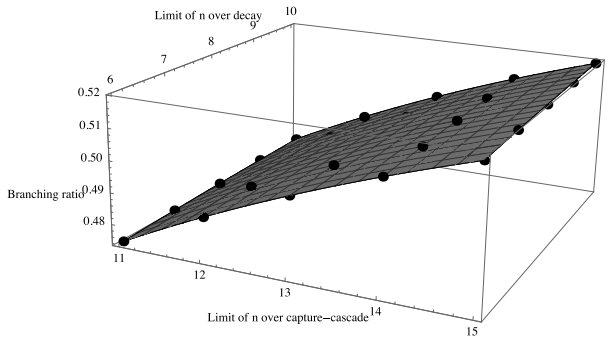}
\caption{The branching ratio of decay into an electron-positron pair from M at a gas temperature of $T=10^{4}$~K as a function of the limiting value of the sums over $n$ in equations~\ref{eqn:fnL} and \ref{eqn:pelpos}, shown by the points.  We find good fits with a  function of the form given in equation~\ref{eqn:ffit2}, shown by the curve; in this case giving a limiting value as $n \rightarrow \infty$ of 0.59.}
\label{fig:eefit}
\end{figure}

For example, Figure~\ref{fig:MLyafit} shows the fit to the branching ratio for M emitting Lyman $\alpha$ radiation at a gas temperature of $T=10^{4}$~K.  We find a good fit to all curves using a function of the form
\begin{equation}
\label{eqn:ffit1}
b = a + c {\rm e}^{- \frac{n}{d}},
\end{equation}
where $a$, $c$, and $d$ are all parameters to be fit.
Figure~\ref{fig:eefit} shows the two dimensional fit over both limits for annihilation into electron-positron pairs for M at $T=10^{4}$~K.  Again, we find a good fit with a functional form 
\begin{equation}
\label{eqn:ffit2}
b = a + c {\rm e}^{- \frac{n_{1}}{d}}+ f {\rm e}^{- \frac{n_{2}}{g}},
\end{equation}
where where $a$, $c$,  $d$, $f$ and $g$ are all parameters to be fit, and the $n_{1}$ and $n_{2}$ are the different limiting values of $n$ in the sums.

We have checked the accuracy of the extrapolations in the following way.  The branching ratios for the two photon or three photon decay of Ps, should be 0.25 and 0.75 by definition, since there is no other way for Ps to decay.  We find the values given in Table~\ref{tab:Psbranch}, which agree to within 2 per cent in all cases.

We wish to highlight one further caveat to these branching ratios.  The extrapolations for photon decays, and electron-positron decays are not large, and become less important as $n$ increases, since they can only occur when $L=0$, which becomes less common for higher $n$.  However, the decay of either particle can occur from any $L$ level, and the extrapolations are large and therefore uncertain.  Therefore, rather than rely on the extrapolations for decay,  we set the probability to be one minus the sum of the other decays.  In any case, the decay of either particle does not have a clear observational signature and it is the probabilities for the other processes which are most important for this study.  This uncertainty in our results could be lessened through calculations to higher levels, but would require significant computing resources which are unavailable to us at this time.

The results for the branching ratios are given in Tables~\ref{tab:Psbranch},~\ref{tab:Mbranch} and \ref{tab:Tbranch}.  Since Ps is well studied and understood, we do not discuss it further\cite{wal96,pra11}.  

M has significant probabilities of Lyman $\alpha$ and Balmer $\alpha$ radiation before decaying.  The decay processes are dominated by two-photon annihilation for para-M and electron-positron annihilation for ortho-M, as well as significant probabilities for either $\mu^{\pm}$ to decay if $T < 10^{5}$~K.  

The much shorter lifetimes of $\tau$ means that the decay processes of T are dominated by the decay of either $\tau^{\pm}$, with only a small probability of either electron-positron or two photon annihilation.  The probabilities of  emitting Lyman $\alpha$ radiation before decay are 1 -- 2 per cent, and $\approx 0$ for other emission lines.

\begin{table*}
\caption{Total branching ratios for Ps as a function of temperature.}
\label{tab:Psbranch}
\begin{tabular}{lllllll}
%T (K) & Ly$\alpha$ &  Ly$\beta$ & Balmer $\alpha$ & Balmer $\beta$ & Two photon decay & Three photon decay  \\ \hline
&&&&& Two & Three   \\ 
&&&&&photon&photon \\
T (K) & Ly$\alpha$ &  Ly$\beta$ & Balmer $\alpha$ & Balmer $\beta$ &decay&decay\\\hline
1000 & 0.46 & 0.08 & 0.24 & 0.06 & 0.24 & 0.73 \\
 10000 & 0.29 & 0.07 & 0.09 & 0.03 & 0.25 & 0.74 \\
 100000 & 0.11 & 0.03 & 0.03 & 0.01 & 0.25 & 0.75 \\
 1000000 & 0.04 & 0.01 & 0.02 & 0.01 & 0.25 & 0.75 \\
 10000000 & 0.03 & 0.01 & 0.02 & 0.01 & 0.25 & 0.75 \\
\end{tabular}
\end{table*}

\begin{table*}
\caption{Total branching ratios for M as a function of temperature.}
\label{tab:Mbranch}
\begin{tabular}{lllllllll}
&&&&&Two&Three&&\\
&&&&&photon&photon&Electron-positron&Decay of\\
T (K) & Ly$\alpha$ &  Ly$\beta$ & Balmer $\alpha$ & Balmer $\beta$ & decay & decay  &decay &either particle\\ \hline
1000 & 0.37 & 0.04 & 0.27 & 0.04 & 0.16 & 0.00 & 0.48 & 0.36\\
 10000 & 0.45 & 0.05 & 0.31 & 0.05 & 0.20 & 0.00 & 0.59 & 0.21\\
 100000 & 0.46 & 0.07 & 0.26 & 0.06 & 0.23 & 0.00 & 0.70 & 0.07 \\
 1000000 & 0.32 & 0.07 & 0.12 & 0.04 & 0.25 & 0.00 & 0.74 & 0.01 \\
 10000000 & 0.13 & 0.04 & 0.02 & 0.01 & 0.25 & 0.00 & 0.74 &  0.00 \\
   \end{tabular}
\end{table*}

\begin{table*}
\caption{Total branching ratios for T as a function of temperature.}
\label{tab:Tbranch}
\begin{tabular}{lllllllll}
&&&&&Two&Three&&\\
&&&&&photon&photon&Electron-positron&Decay of\\
T (K) & Ly$\alpha$ &  Ly$\beta$ & Balmer $\alpha$ & Balmer $\beta$ & decay & decay  &decay &either particle\\ \hline
 1000 & 0.01 & 0.00 & 0.00 & 0.00 & 0.03 & 0.00 & 0.06 & 0.91 \\
 10000 & 0.01 & 0.00 & 0.00 & 0.00 & 0.03 & 0.00   & 0.07 & 0.90 \\
 100000 & 0.01 & 0.00 & 0.00 & 0.00 & 0.04 & 0.00   & 0.09 & 0.87 \\
 1000000 & 0.01 & 0.00 & 0.00 & 0.00 & 0.06 &   0.00 & 0.12 & 0.83 \\
 10000000 & 0.02 & 0.00 & 0.00 & 0.00 & 0.08 &   0.00 & 0.17 & 0.75 \\
   \end{tabular}
\end{table*}

\section{Observable signatures}
\label{sec:signatures}

Tables~\ref{tab:Mbranch} and~\ref{tab:Tbranch} give the branching ratios for possible radiative recombination lines and decays of M and T.  We now discuss  the observational signatures of these processes.   
The energies of radiative recombination lines and of 2-photon annihilation signatures and ionization are given in Table~\ref{tab:summ}.  In addition to the 2-photon annihilation signature there will be a continuum of radiation from the 3-photon decay of the singlet state, with energies from zero to the energy of the 2-photon decay.  The decay mechanisms for both $\tau$ and $\mu$ are numerous and there is no one clear observable signature that we can associate with either of these, nor is there any single identifiable signature of annihilation into e$^{\pm}$.

%\begin{table*}
%\caption{Energies of radiative recombinations, 2-photon annihilation signatures and ionization energies..}
%\label{tab:wavelengths}
%%\begin{tabular}{lllllllllll}
%%& \multicolumn{4}{c}{Wavelength (\AA)} &  \multicolumn{4}{c}{Energy (keV)} & \multicolumn{2}{c}{Energy} \\
%%& Lyman $\alpha$ & Lyman $\beta$ & Balmer $\alpha$ & Balmer $\beta$ & Lyman $\alpha$ & Lyman $\beta$ & Balmer $\alpha$ & Balmer $\beta$ & Annihilation (MeV) & Ionization (eV)\\ \hline
%%Ps & 2430.0 & 2050.4 & 13122 & 9720.2 &$5.102 \times 10^{-3}$ &$6.047 \times 10^{-3}$ &$9.448 \times 10^{-4}$&$1.276 \times 10^{-3}$&0.51100&6.8\\
%%M &11.753 & 9.9162 & 63.464 & 47.010 &1.055&1.250&0.195&0.264&105.66&1406.6\\
% %T& 0.70 & 0.59&  3.8 & 2.8 &17.7&21.0&3.3&4.4&1784.1&23751.4 
 %\begin{tabular}{lllllll}
%& \multicolumn{4}{c}{Radiative recombination (keV)} &\\
%&  Lyman $\alpha$ & Lyman $\beta$ & Balmer $\alpha$ & Balmer $\beta$ & Annihilation (MeV) & Ionization (eV)\\ \hline
%%Ps & 2430.0 & 2050.4 & 13122 & 9720.2 &$5.102 \times 10^{-3}$ &$6.047 \times 10^{-3}$ &$9.448 \times 10^{-4}$&$1.276 \times 10^{-3}$&0.51100&6.8\\
%M &1.055&1.250&0.195&0.264&105.66&1406.6\\
 %T& 17.7&21.0&3.3&4.4&1784.1&23751.4 
%\end{tabular}
%\end{table*}

All these signatures are observable with current instrumentation if they are sufficiently bright.  For example Fermi-LAT is sensitive to $\gamma$-rays between 30 MeV to 300GeV and could observe the annihilation of M and T; 
%INTEGRAL SPI  can (and does) observe the annihilation of Ps;  
INTEGRAL IBIS is sensitive to the T Lyman lines; 
XMM-Newton is sensitive to X-rays between 0.1 and 15 keV, and can observe the recombination lines of M, and the T Balmer lines.
%; HST COS is sensitive to the Lyman recombination lines of Ps; the Balmer recombination line of Ps is observable with a large number of ground based near-infrared spectrographs. 
 
\section{Expected signatures from astrophysical sources}
\label{sec:astrophys}

We are finally in a position to estimate the expected observable signatures of M and T from astrophysical sources.
Consider a source of particle anti-particle pairs being produced at a rate $r$ per second.  
The fraction of these pairs which will produce M or T is given by
\begin{equation}
f_{\rm onium} = r f_{\rm ion} g_{\rm rr}
\end{equation}
where
$f_{\rm ion}$ is the fraction of pairs produced with energy less than the ionization energy, i.e.\ $f_{\gamma \gamma}$ or $f_{ee}$ in sections~\ref{sec:gg} and \ref{sec:ee} above, and
$g_{\rm rr}$ is the fraction of those pairs which radiatively recombine (\S~\ref{sec:gould}, Figure~\ref{fig:rrda}).  Therefore the luminosity of any particular observable signature in ph~s$^{-1}$ is 
\begin{equation}
\label{eqn:lum}
L = p\ f_{\rm onium}\ b 
\end{equation}
where $b$ is the branching ratio of the particular signature (Tables~\ref{tab:Mbranch} and \ref{tab:Tbranch}), and $p$ is a factor which specifies the number of photons produced by the process, i.e.\ $p=1$ for recombination lines, $p=2$ for two-photon annihilation and $p=3$ for three-photon annihilation.  These luminosities can be easily converted to units of erg~s$^{-1}$ using the energies of the emitted photons given in Table~\ref{tab:summ}.

We will now estimate the expected fluxes from particular sources.  These should be compared to the limiting sensitivities of current instruments.  The M recombination lines are all in the soft X-ray band (0.5 -- 2 keV), for which a 100 ksec observation has a $\approx 4\sigma$ limiting sensitivity\cite{has01} of $f_{\rm X}\sim 3.1 \times 10^{-16}$ erg cm$^{-2}$ s$^{-1}$.  The T recombination Balmer lines are  in the hard X-ray band (2 -- 10 keV), for which a 100 ksec observation has a $\approx 4\sigma$ limiting sensitivity of $f_{\rm X}\sim 1.4 \times 10^{-15}$ erg cm$^{-2}$ s$^{-1}$.    The T Lyman lines  are at the low energy limit of the SPI spectrometer  on board INTEGRAL.  The limiting 4$\sigma$ sensitivity for an exposure time of $10^{5}$ s is $\approx 3 \times 10^{-13}$ erg cm$^{-2}$ s$^{-1}$, using data from the INTEGRAL webpages.  The M and T annihilation lines are detectable by FERMI-LAT, for which the limiting $4\sigma$ sensitivity for an exposure time of $10^{5}$ s is $\approx 4 \times 10^{-13}$ erg cm$^{-2}$ s$^{-1}$ for M and $\approx 1 \times 10^{-13}$ erg cm$^{-2}$ s$^{-1}$ for T, using data from the FERMI webpages.

\subsection{Jets}

The powerful relativistic jets of active galactic nuclei (AGN) and microquasars are a probable source of pair production.  These jets have strong radio emission with a power-law spectrum which is interpreted as being due to synchrotron emission from relativistic $e^{-}$ gyrating around the magnetic field lines of the AGN.  The jets must be electrically neutral, otherwise they would cause a potential difference to build-up, which would oppose and eventually  stop the jet.  

It is unknown whether the positive component of jets consists of positrons, protons, or a mixture of both.  Observational studies have had to rely on indirect methods of searching for the presence of \e+, such as estimates of the bulk kinetic energy contained in jets, which have been used to argue for both $e^{-}$-p plasma\cite{cel93} and $e^{-}$-\e+\ plasma\cite{rey96,war98,hir05}.   

Theoretically there are good reasons to expect that jets contain some fraction of \e+.  A pair plasma has the advantage of explaining $\gamma$-ray jets\cite{bla95} and the very high Lorentz factors ($\Gamma >  5$) required to account for superluminal bulk velocities of jets\cite{beg84}.

It is possible that some Ps may form in the jet, or similarly, that some e$^{\pm}$ pairs may collide with the required energy to form M and T as outlined in \S~\ref{sec:ee}.  However, in the observers frame, the velocity of the oniums thus formed will be highly relativistic, causing relativistic Doppler broadening and beaming of any emitted radiation\cite{boe96,mar07}.  Such a broadened signal may still be possible to observe\cite{boe96}, but would be difficult to interpret, and hardly constitutes an unambiguous test of the presence of exotic onium atoms such as M or T.
Therefore we do not consider onium formation in the jet itself as a likely candidate for detection.

However, if a jet collides with stationary object such as a gas cloud, or a star, then the leptonium formed in the collision will give rise to emission that is in principle observable.   Such collisions may occur in the radio jets of AGN\cite{gom00}, or if the jets of microquasars are mis-aligned and hit the secondary companion\cite{gue06}.

\begin{figure}
\begin{center}
\subfigure[M]{
\includegraphics[scale=0.6]{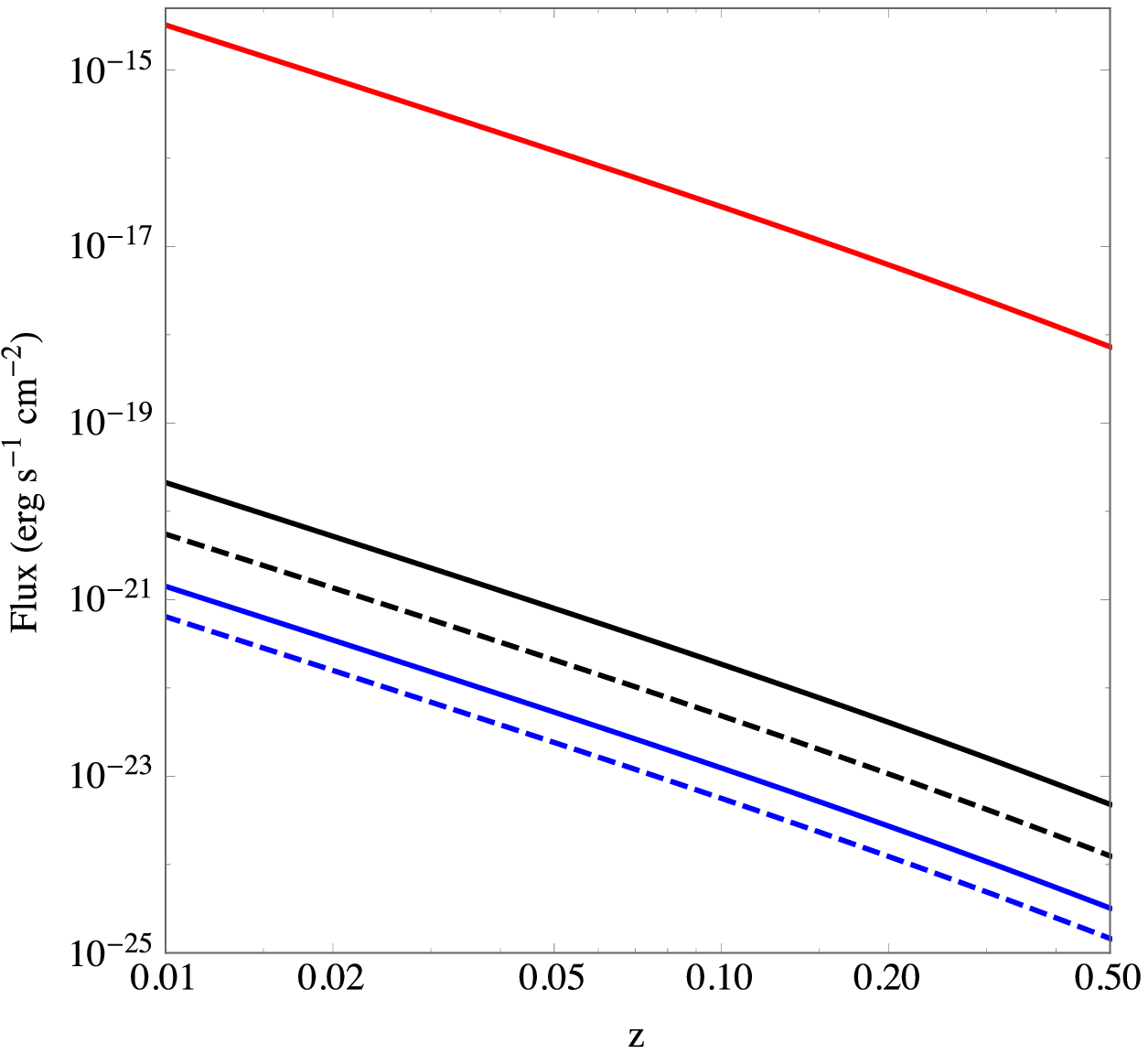}
}
\subfigure[T]{
\includegraphics[scale=0.6]{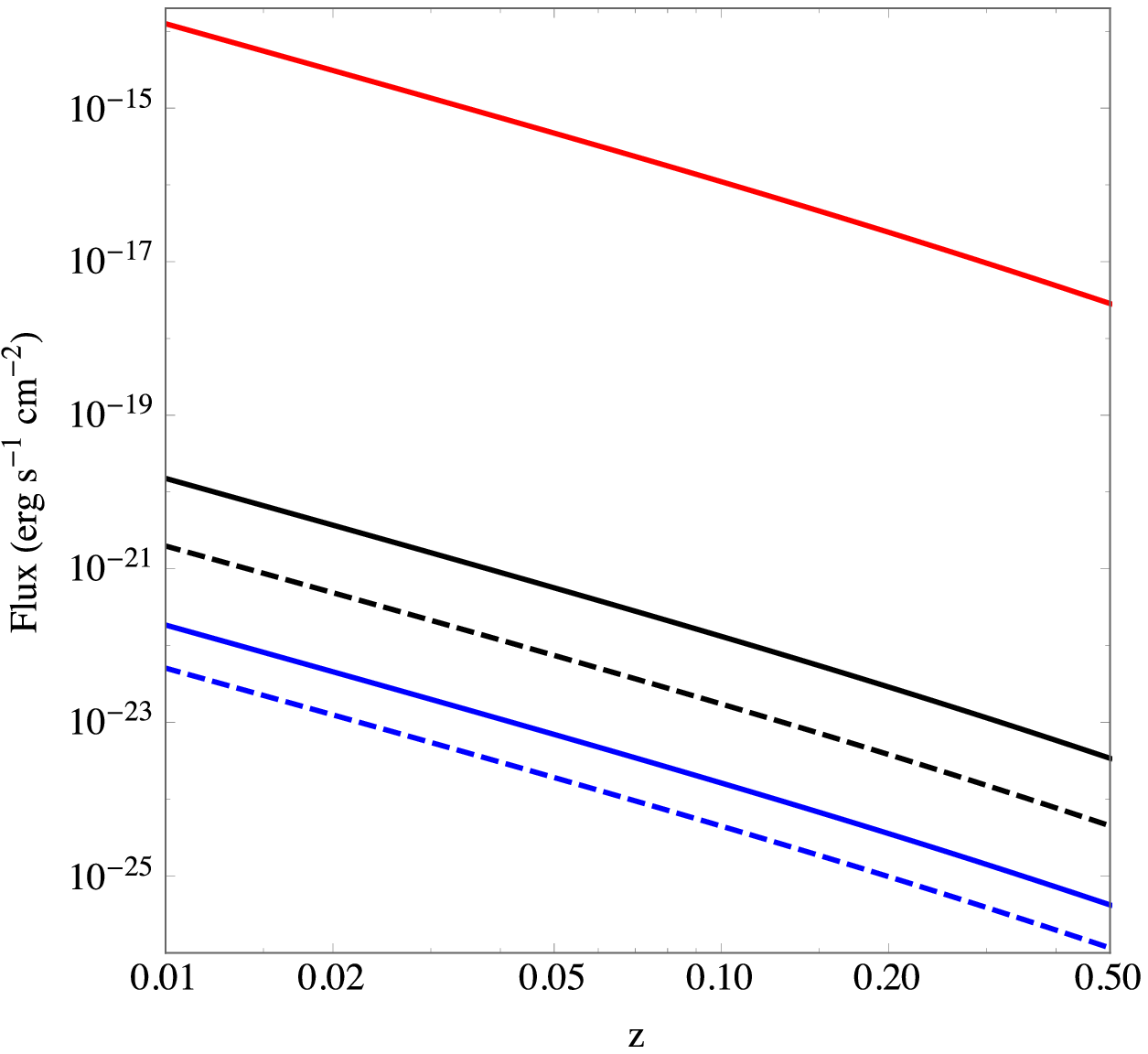}
}
\caption{The flux as a function of redshift for a blazar jet which produces $10^{49}$~\e+~s$^{-1}$, with spectral index $\alpha=1.5$ and $T=10^{6}$~K, colliding with a gas cloud.   The lines show two-photon annihilation (red), Ly $\alpha$ (black) Ly $\beta$ (dashed, black), Ba $\alpha$ (blue) and Ba $\beta$ (dashed, blue).}
\label{fig:blazar}
\end{center}\end{figure}

\begin{figure}
\begin{center}
\subfigure[M]{
\includegraphics[scale=0.6]{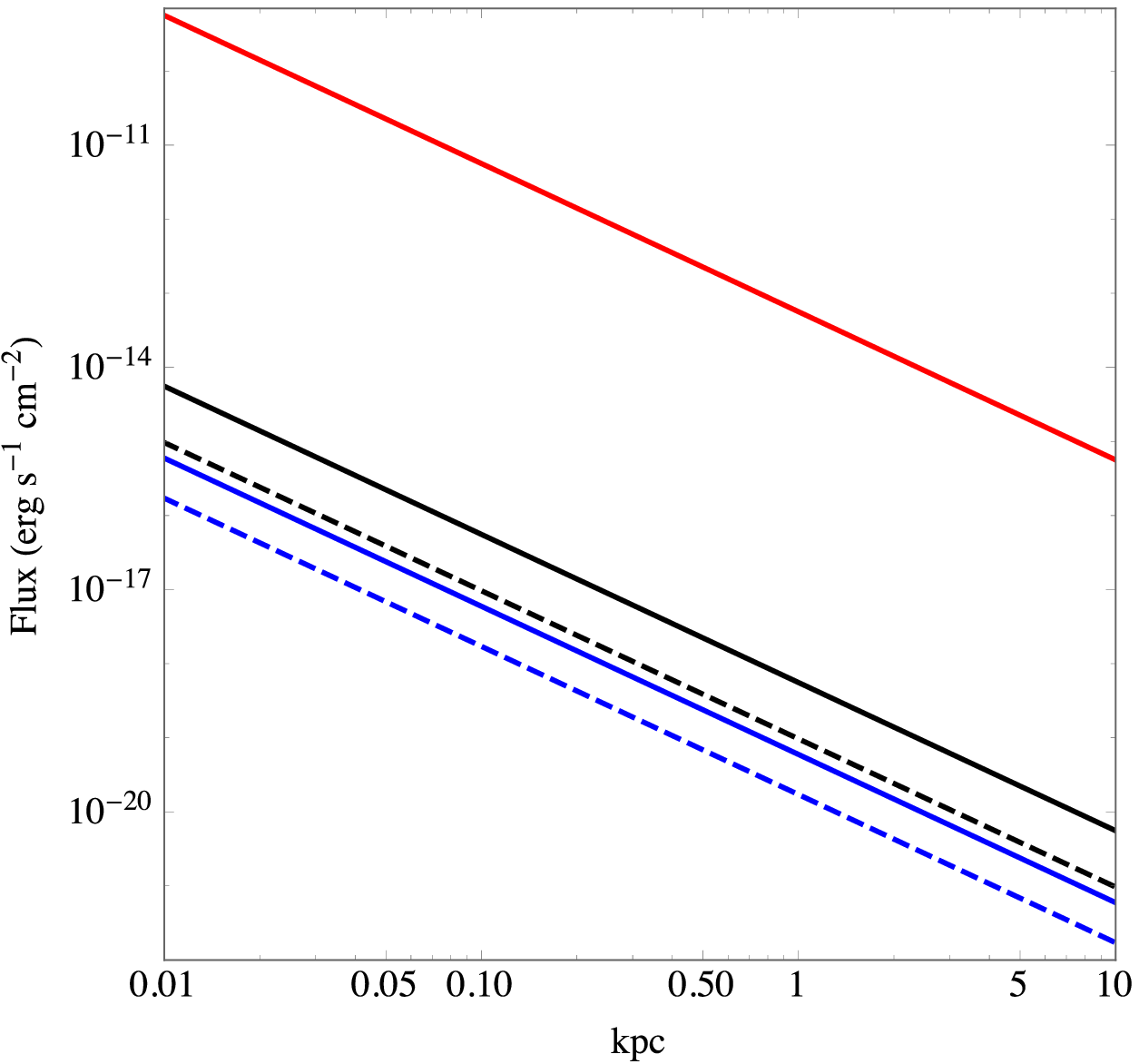}
}
\subfigure[T]{
\includegraphics[scale=0.6]{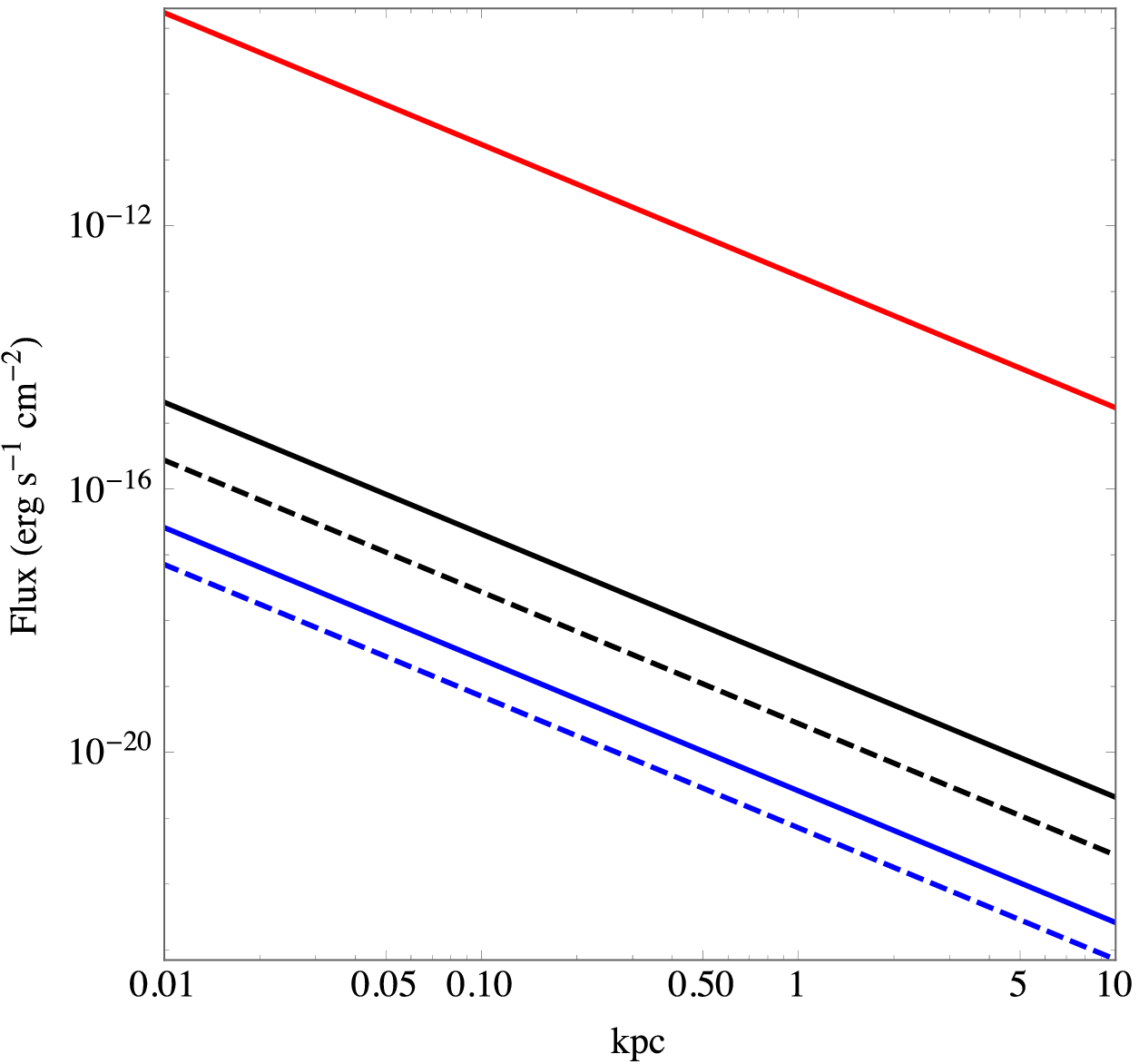}
}
\caption{The flux as a function of redshift for a microquasar jet which produces $10^{41}$~\e+~s$^{-1}$, with spectral index$=-1.5$ and $T=10^{5}$~K, colliding with a gas cloud.  The lines show two-photon annihilation (red), Ly $\alpha$ (black) Ly $\beta$ (dashed, black), Ba $\alpha$ (blue) and Ba $\beta$ (dashed, blue).}
\label{fig:mqso}
\end{center}\end{figure}

Let us consider some illustrative examples.  Previously\cite{ell09}, we have calculated the expected positron contents of jets, by applying  the  arguments of \citet{mar07}  and \citet{mar83} , which were developed in order to search for 511 keV Ps annihilation radiation from 3C 120, to the empirical measurements of \citet{ghi93}. 
We found a maximum of $\approx 10^{49}$~\e+~s$^{-1}$ are produced in the jets of blazars, whereas quasars produce $\approx 10^{46}$ -- $10^{48}$~\e+~s$^{-1}$ in their jets.  The jets of microquasars are expected to have a positron flux of $\approx 10^{41}$~\e+~s$^{-1}$\cite{gue06}.
We assume that the object being hit by the jet is dense enough to stop \emph{all} these positrons.  Let us assume a spectral index of $\alpha = 1.5$ and a temperature of $T=10^{6}$~K for the positrons in the jet.  Thus the intrinsic luminosities of the various observational signatures can be calculated, and are given in Table~\ref{tab:blazar} for a blazar.  These are then used to calculate the flux of the signal as a function of redshift, as shown in Figure~\ref{fig:blazar}.  Similar results are shown in Table~\ref{tab:mqso} and Figure~\ref{fig:mqso}, for a mis-aligned microquasar with a spectral index of $\alpha = 1.5$ and a temperature of $T=10^{5}$~K for the positrons in the jet.  

\begin{table}
\caption{The intrinsic luminosities for a blazar jet which produces $10^{49}$~\e+~s$^{-1}$, with spectral index $ \alpha=1.5$ and $T=10^{6}$~K.}
\label{tab:blazar}
\begin{tabular}{lll}
& \multicolumn{2}{c}{Luminosity (ph s$^{-1}$) }\\
& M & T \\ \hline
Ly $\alpha$ &$2.7\times 10^{42}$ &$1.2\times   10^{41}$\\
Ly $\beta$ & $6.0\times 10^{41}$&$1.3\times   10^{40}$\\
Ba $\alpha$ & $9.8\times 10^{41}$&$7.7\times   10^{39}$\\
   Ba $\beta$ & $3.3\times 10^{41}$&$1.6\times   10^{39}$\\
Two photon & $4.1\times   10^{42}$&$9.6\times   10^{41}$
\end{tabular}
\end{table}

\begin{table}
\caption{The intrinsic luminosities for a microquasar jet which produces $10^{41}$~\e+~s$^{-1}$, with spectral index $\alpha=1.5$ and $T=10^{5}$~K.}
\label{tab:mqso}
\begin{tabular}{lll}
& \multicolumn{2}{c}{Luminosity (ph s$^{-1}$) }\\
& M & T \\ \hline
Ly $\alpha$ &$3.9\times 10^{34}$ &$8.7\times   10^{32}$\\
Ly $\beta$ & $5.8\times 10^{33}$&$9.7\times   10^{31}$\\
Ba $\alpha$ & $2.3\times 10^{34}$&$5.9\times   10^{31}$\\
   Ba $\beta$ & $4.8\times 10^{33}$&$1.2\times   10^{32}$\\
Two photon & $4.0\times   10^{34}$&$7.2\times   10^{33}$
\end{tabular}
\end{table}

For our illustrative examples, no lines from blazars are above the relevant current 4$\sigma$ detection limits for a 100~ks exposure for either M or T.  However,  a microquasar jet which produces $10^{41}$~\e+~s$^{-1}$, with spectral index$=-1.5$ and $T=10^{5}$~K, colliding with a gas cloud, would be detectable in principle to the distances given in Table~\ref{tab:mqsojet}.  The recombination lines would only be detectable at very close distances, but the two photon annihilation should be detectable at 4$\sigma$ in a 100~ksec exposure to $\approx 0.4$~kpc and $\approx 1$~kpc  for M and T respectively.

\begin{table}
\caption{The distance to which a microquasar jet which produces $10^{41}$~\e+~s$^{-1}$, with spectral index$=-1.5$ and $T=10^{5}$~K, colliding with a gas cloud, would be detectable at 4 $\sigma$ in 100 ksec.}
\label{tab:mqsojet}
\begin{tabular}{lcc}
& \multicolumn{2}{c}{Distance (kpc)} \\
&M&T\\ \hline
Ly $\alpha$ &0.04 & $9 \times 10^{-4}$\\
Ly $\beta$ & 0.01 & $3 \times 10^{-4}$\\
Ba $\alpha$ & 0.01 & 0.001\\
   Ba $\beta$ &0.007 &$ 7 \times 10^{-4}$\\
Two photon & 0.40 & 1.13
\end{tabular}
\end{table}

\subsection{Accretion discs}

We now consider pairs produced in the accretion disc itself through photon-photon annihilation.    The number density of pairs thus produced in the optically thick region of the disc is
\begin{equation}
N_{\gamma\gamma} \sim \frac{1}{\sigma_{\rm T} R},
\end{equation}
where $\sigma_{\rm T}$ is the Thompson cross-section for $\mu$ or $\tau$ and $R$ is the radius of the optically thick disc\cite{belo99}.  If we assume $R \approx 2GM/c^{2}$, i.e.\ the disc has approximately the Schwarzschild radius, and the thickness of the disc is $\approx R$, then taking masses of $M_{\rm AGN} = 10^{6}$ M$_{\odot}$ for the mass of a black hole in an AGN, and  $M_{\mu{\rm QSO}} = 10$ M$_{\odot}$ for the mass of a black hole in a microquasar, then the pair yields are as given in Table~\ref{tab:accdisc}.

\begin{table}
\caption{The number of pairs produced through photon-photon annihilation in an accretion disc.}
\label{tab:accdisc}
\begin{tabular}{lll}
& AGN & Microquasar \\ \hline
$\mu$ & $1.7 \times 10^{52}$ & $1.7 \times 10^{42}$ \\
$\tau$ & $5.0 \times 10^{54}$ & $5.0 \times 10^{44}$ \\
\end{tabular}
\end{table}

\begin{figure}
\begin{center}
\subfigure[M]{
\includegraphics[scale=0.6]{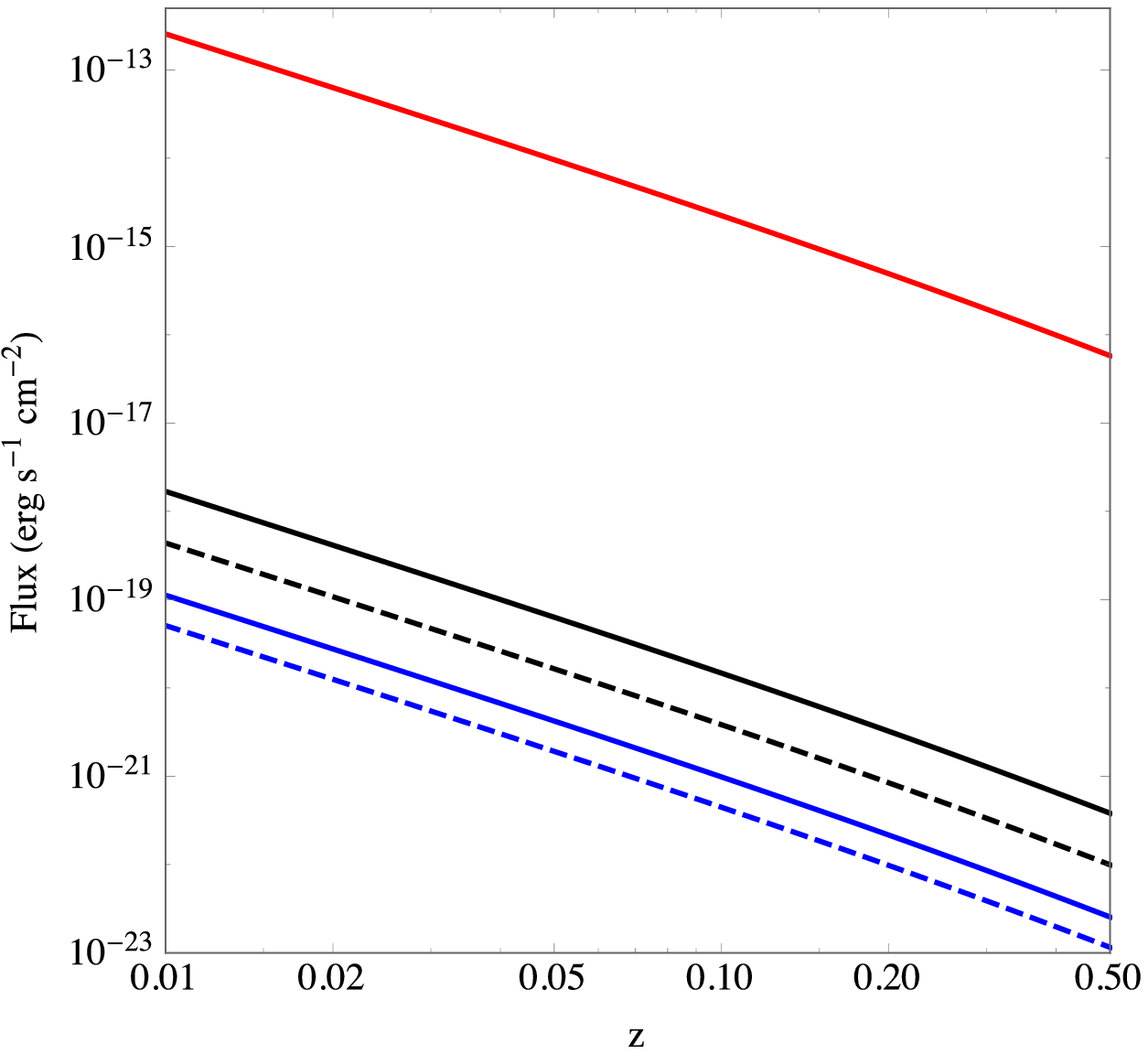}
}
\subfigure[T]{
\includegraphics[scale=0.6]{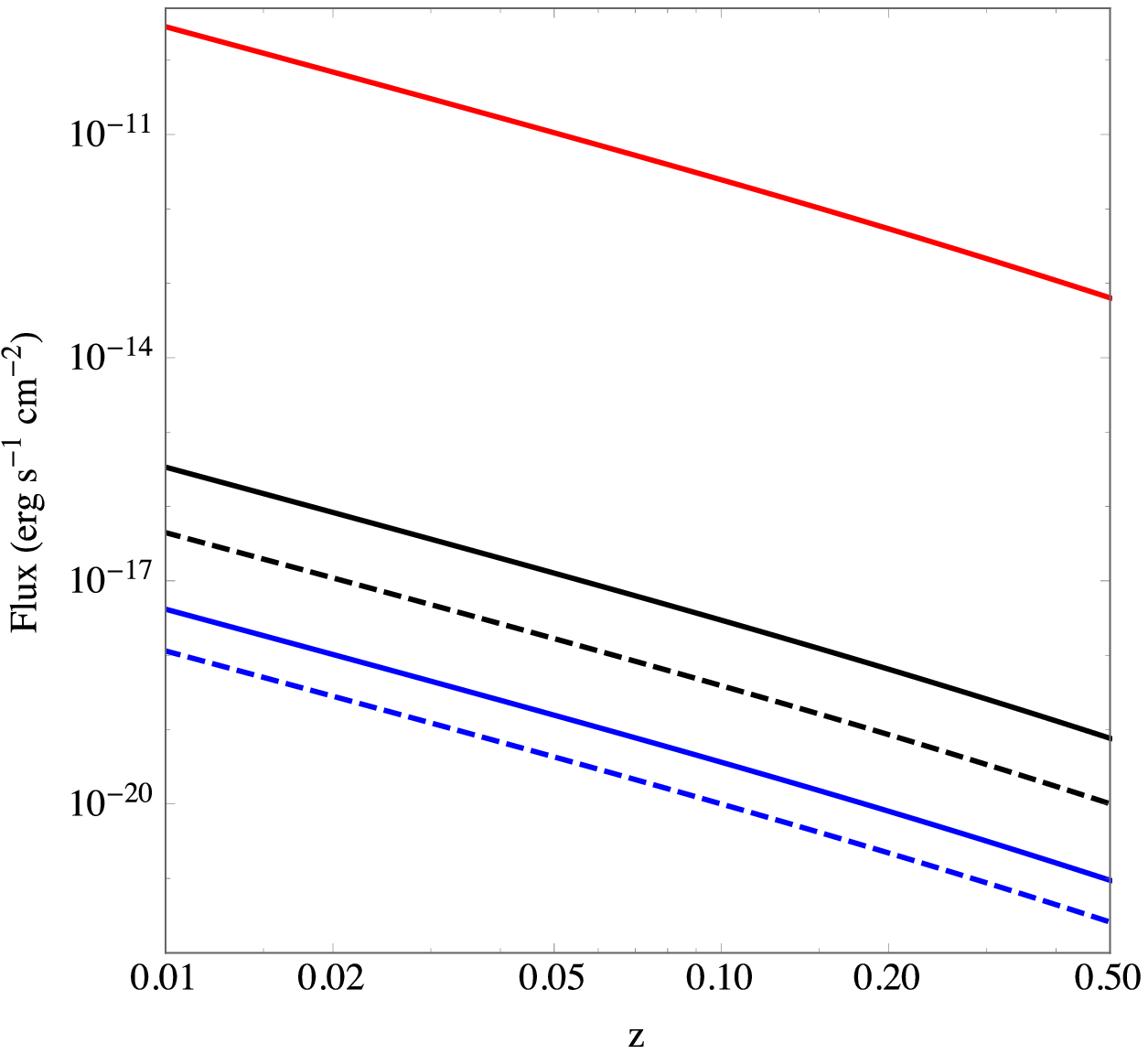}
}
\caption{The flux as a function of redshift for an accretion disc around an AGN with a $10^{6}$~M$_{\odot}$ black hole, and spectral index $\alpha=1.5$ and $T=10^{6}$~K.  The lines show two-photon annihilation (red), Ly $\alpha$ (black) Ly $\beta$ (dashed, black), Ba $\alpha$ (blue) and Ba $\beta$ (dashed, blue).}
\label{fig:accdiscagn}
\end{center}\end{figure}

\begin{figure}
\begin{center}
\subfigure[M]{
\includegraphics[scale=0.6]{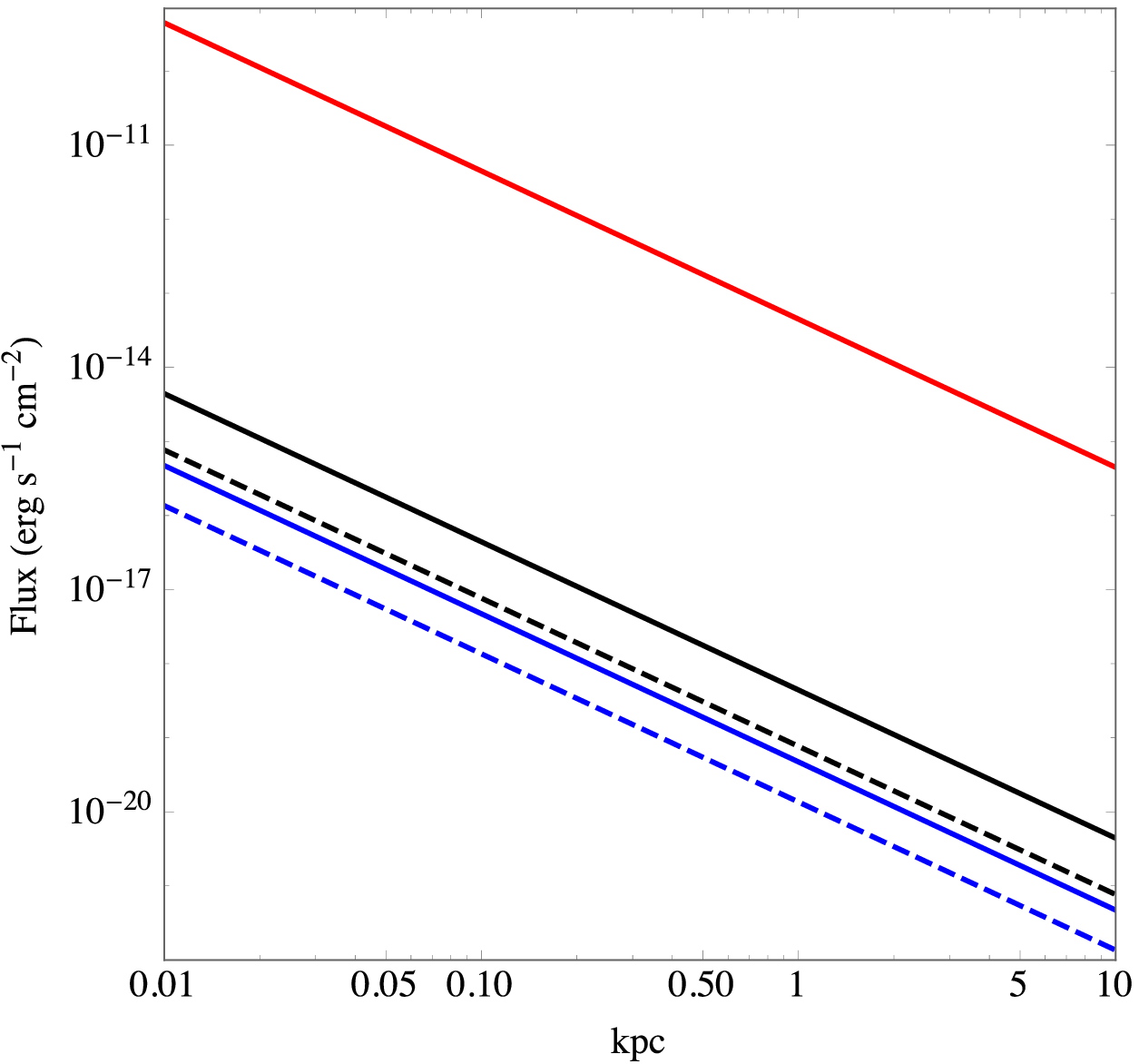}
}
\subfigure[T]{
\includegraphics[scale=0.6]{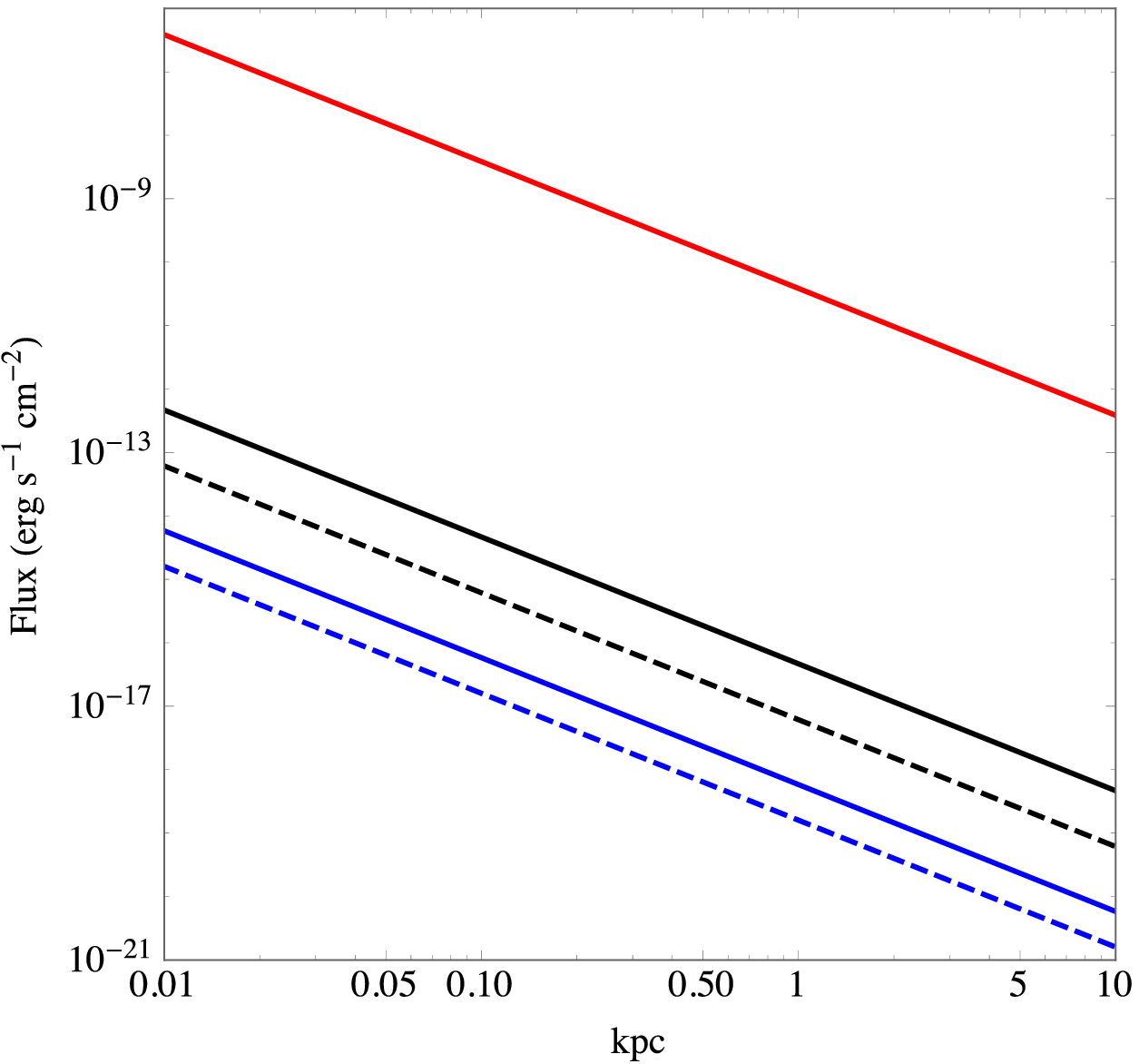}
}
\caption{The flux as a function of redshift for an accretion disc around a microquasar with a $10$~M$_{\odot}$ black hole, and spectral index $\alpha=1.5$ and $T=10^{5}$~K.  The lines show two-photon annihilation (red), Ly $\alpha$ (black) Ly $\beta$ (dashed, black), Ba $\alpha$ (blue) and Ba $\beta$ (dashed, blue).}
\label{fig:accdiscmqso}
\end{center}\end{figure}

Now, assuming a spectral index of $\alpha = 1.5$, then we may estimate the branching ratios as in equation~\ref{eqn:lum}, whereupon we find the luminosities in ph~s$^{-1}$, as listed in Table~\ref{tab:accdisclum}.  The flux of the AGN as a function of redshift, and the flux of the microquasar as a function of distance, are shown in Figures~\ref{fig:accdiscagn} and Figures~\ref{fig:accdiscmqso}.

\begin{table}
\caption{The intrinsic luminosities for pair production via photon annihilation in accretion discs surrounding a microquasar and an AGN, assuming solar masses of $10$ and $10^{6}$ M$_{\odot}$ respectively, temperatures of $10^{5}$~K and $10^{6}$~K, and a spectral index of $\alpha=1.5$.}
\label{tab:accdisclum}
\begin{tabular}{lllll}
& \multicolumn{2}{c}{Microquasar} & \multicolumn{2}{c}{AGN} \\
& \multicolumn{2}{c}{Luminosity (ph s$^{-1}$) } & \multicolumn{2}{c}{Luminosity (ph s$^{-1}$) } \\
& M & T &M&T\\ \hline
Ly $\alpha$ & $2.9 \times 10^{35}$  &$1.8 \times 10^{36}$      & $2.0 \times 10^{45}$     & $2.4 \times 10^{46}$\\
Ly $\beta$   & $4.2 \times 10^{34}$  &$2.0 \times 10^{35}$         &$4.4 \times 10^{44}$      &  $2.7 \times 10^{45}$\\
Ba $\alpha$ &   $1.7 \times 10^{35}$  &  $1.2 \times 10^{35}$  &  $7.2 \times 10^{44}$     & $1.6 \times 10^{45}$ \\
 Ba $\beta$  & $3.6 \times 10^{34}$  &$2.5 \times 10^{34}$        &  $2.4 \times 10^{44}$     &   $3.2 \times 10^{44}$\\
Two photon & $2.9 \times 10^{35}$  & $1.5 \times 10^{37}$      &  $3.0 \times 10^{45}$     &  $2.0 \times 10^{47}$
\end{tabular}
\end{table}

For these illustrative examples, and a  4$\sigma$ detection limit for a 100~ks exposure,  the two-photon annihilation line of M would be observable in the accretion disc of an AGN at $z<0.025$, while the two-photon annihilation line of T would be observable out to $z<0.9$.  Neither the M nor the T recombination lines would be detectable.

Meanwhile, for microquasars the M and T lines would be detectable at 4$\sigma$ in a 100 ksec exposure to the distances given in Table~\ref{tab:mqsodisc}.  The recombination lines are detectable at close distances while the two photon annihilation should be detectable to 1.1 or 51~kpc for M and T respectively.  

\begin{table}
\caption{The distance to which  an accretion disc around a microquasar with a $10$~M$_{\odot}$ black hole, and spectral index $\alpha=1.5$ and $T=10^{5}$~K, would be detectable at 4 $\sigma$ in 100 ksec.}
\label{tab:mqsodisc}
\begin{tabular}{lcc}
& \multicolumn{2}{c}{Distance (kpc)} \\
&M&T\\ \hline
Ly $\alpha$ & 0.11&0.04 \\
Ly $\beta$ & 0.05& 0.01\\
Ba $\alpha$ & 0.04 & 0.06 \\
   Ba $\beta$ &0.02 & 0.03 \\
Two photon &1.1 & 51
\end{tabular}
\end{table}

\section{Conclusions}
\label{sec:discuss}

True muonium and true tauonium are the most compact pure QED systems, but have never been observed.  Unlike Ps, for which there are extensive observations in our own Galaxy\cite{pra11}, both the formation and decay of M and T are  affected by the intrinsic instability of the $\mu$ and $\tau$ leptons.  We have investigated the likelihood of their formation in astrophysical environments, and the prospects for their observation.  

The probability of formation is small, $\sim 10^{-7}$ from photon-photon annihilation or electron-positron annihilation.  The probability is small for two reasons: (i) the lifetimes of the $\mu$ and $\tau$ are intrinsically short (\S~\ref{sec:mutaudecay}), and thus M and T can only form from the products of pair production processes, such that the pairs immediately recombine, (ii) even then, only those pairs with a total kinetic energy less than the ionisation energy can form leptonium, and since pair production usually takes place in high energy process, these pairs constitute a small fraction of the total.
Nevertheless,  high energy astrophysical environments are capable of producing copious numbers of $\mu$ and $\tau$ pairs, and the cross-section for radiative recombination dominates that of direct annihilation (\S~\ref{sec:gould}).  Thus even the small fraction of pairs with energy low enough to
 form M or T can lead to a significant flux.  

The decay of M and T are  hastened by the short lifetimes of the $\mu$ and $\tau$ leptons, which may be one reason why the possibility of  astrophysical  observations has not received much attention.    Here, we have carefully calculated the probabilities of the various observational signatures.  We have calculated the cross-sections for recombination onto the $nL$th level (\S~\ref{sec:wallyn}), and the resulting branching ratios for Lyman $\alpha$, Lyman $\beta$, Balmer $\alpha$, Balmer $\beta$, two photon annihilation, three photon annihilation, annihilation into e$^{\pm}$ pairs, and the decay of either $\mu$ or $\tau$.  Although the decay of either $\tau$ does dominate for T, there is still a small probability for observing recombination lines or two-photon annihilation.  For M the situation is more hopeful, with significant branching ratios for Lyman $\alpha$, Balmer $\alpha$ and two-photon annihilation.

In section~\ref{sec:astrophys}, we made estimates of the fluxes of M and T recombination and annihilation signatures for the cases of blazar jet-cloud interactions, jet-star interactions in mis-aligned microquasars, and within the accretion discs of AGN and microquasars.  These were compared to the current detection limits of X-ray and $\gamma$-ray observatories.  The expected signatures from AGN jet-cloud interactions were all below current detections limits.  
However,   M and T formation within microquasar jet-star interactions, or within the accretion discs of both AGN and microquasars  were estimated to yield signatures brighter than the detection limits, with those from microquasars offering the brightest estimates due to their proximity.  Actual observations would be further complicated by the intrinsic backgrounds and also emission from the object in question, which may have other significant line emission (e.g.\ see \citet{mar02} for a spectrum of SS 433).

These examples are only illustrative.   Other sources may also be significant.   For example, 511 keV electron-positron annihilation radiation was recently discovered in  terrestrial $\gamma$-ray flashes due to lightning strikes\cite{bri11}.  These same flashes could lead to M or T formation, as well as the possibility of observing Ps recombination lines.  
As a different example we note in passing that the recent detection of an unidentified line at $\approx 3.5$~keV in galaxy clusters\cite{boy14,bul14} cannot be explained by the T Balmer $\alpha$ line at 3.3 keV, which is ruled out by the constraints on the energy.  % However, such anomlous signatures are routinely discovered, and often lead to speculations of exotic physics

In summary, the astrophysics of pair-production in any high energy source could lead to possible M and T formation.  This paper provides the tools to estimate the fluxes of the M and T recombination and annihilation signatures once the rate of pair production is known.

%\section{Acknowledgments}
\begin{acknowledgments}
We thank the referees for useful comments which have improved this paper.  We thank M.~Voloshin for useful advice regarding the Sommerfeld-Sakharov correction.  We thank M.~Colless for advice on the nomenclature for leptonium which improved the lucidity of the paper.
\end{acknowledgments}

\begin{widetext}
\appendix*

\section{Energy threshold in the laboratory frame for muon pair production from electron-positron annihilation}
\label{app:eemmenergy}

To relate the lab frame energies of the colliding electrons, $E_{1}$ and $E_{2}$, to the zero momentum frame Lorentz factors, $\gamma$, of the produced muons, we note that the absolute square of the relativistic four momentum $p^{2} = p^{\mu} p_{\mu}$ is invariant.  In the lab frame, let the incoming electron be traveling along the x-axis in the positive direction, with speed $u$, and the incoming positron be at an angle $\theta$ to the x-axis, in the x-y plane, with speed $w$.  The relative speed of the particles in the lab frame is then, $v=u - w\cos \theta$.  Therefore in the lab frame the total relativistic four-momentum of both particles is given by,
 \begin{equation}
 p_{\rm lab} = \left(\frac{E_{1}+E_{2}}{c}, p_{1}+p_{2} \cos \theta, p_{2} \sin \theta,0\right).
 \end{equation}
 In the zero-momentum frame, after the collision, the produced muons have equal and opposite momentum, thus,
 \begin{equation}
 p_{\rm z.m.} = \left(\frac{E_{3}+E_{4}}{c},0,0,0\right).
 \end{equation}
 Now since $p^{2}$ is invariant we have,
 \begin{eqnarray}
 p_{\rm lab}^{2} &=& p_{\rm z.m.}^{2} \nonumber \\
 \left(\frac{E_{1}+E_{2}}{c}\right)^{2} - \left(p_{1}+p_{2} \cos \theta\right)^{2} - p_{2}^{2} \sin^{2} \theta
 &=& \left(\frac{E_{3}+E_{4}}{c}\right)^{2} \nonumber \\
E_{1} E_{2} + c^{4} (m^{2} - 2 M^{2} \gamma^{2}) & = & \sqrt{E_{1}^{2} - c^{4} m^{2}} \sqrt{E_{2}^{2} - c^{4} m^{2}} \cos\theta. \label{eqn:eemmen}
 \end{eqnarray}
 %
% And therefore,
 %\begin{equation}
 %E_{1} =  (c^{4} E_{2} (m^2 - 2 M^2 \[Gamma]^2) - 
 %c^2 Abs[Cos[\[Theta]]] Sqrt[(E2^2 - c^4 m^2) (-E2^2 m^2 + 
   %  c^4 (m^2 - 2 M^2 \[Gamma]^2)^2 + 
    % m^2 (E2^2 - c^4 m^2) Cos[\[Theta]]^2)])/(-E2^2 + (E2^2 - 
    %c^4 m^2) Cos[\[Theta]]^2)
Therefore the relative velocity is,
\begin{equation}
v=\frac{\sqrt{c^{2} E_{1}^{2} -  m^{2} c^{6}}}{E_{1}} - \frac{
 \sqrt{c^{2} E_{2}^{2} - m^{2} c^{6} } \cos \theta}{E_{2}} \label{eqn:eemmv}.
 \end{equation}

\end{widetext}

\bibliography{ps}% Produces the bibliography via BibTeX.

\end{document}